\documentclass[aps,pre,
twocolumn,longbibliography,superscriptaddress]{revtex4-2}
\usepackage{bm}
\usepackage{graphicx}
\usepackage{graphics}
\usepackage{amsmath}
\usepackage{amsfonts}
\usepackage{amssymb}
\usepackage{makeidx}
\usepackage{hyperref}
\hypersetup{colorlinks=true,citecolor=blue,linkcolor=blue,filecolor=blue,urlcolor=blue,pdfstartview=FitH,pdfpagemode=UseNone}

\usepackage{color,soul} 
\usepackage{float}
\usepackage{graphicx}
\usepackage{caption}
\usepackage{subcaption}
\usepackage{caption}
\usepackage{graphicx}
\usepackage{ragged2e} 
\usepackage{graphicx}
\usepackage{subcaption}
\usepackage{float} 
\usepackage{xcolor}
\usepackage{tikz}
\usepackage{ragged2e}
\usepackage{physics} 
\usepackage[utf8]{inputenc}
\makeatletter
\long\def\@makecaption#1#2{%
  \vskip\abovecaptionskip
  \begingroup
  \justifying
  \small
  \sbox\@tempboxa{#1: #2}%
  \ifdim \wd\@tempboxa >\hsize
    #1: #2\par
  \else
    \hbox to\hsize{\hfil \box\@tempboxa \hfil}%
  \fi
  \endgroup
  \vskip\belowcaptionskip
}
\makeatother

\usepackage{siunitx}
\usepackage[normalem]{ulem}
\definecolor{violetae}{RGB}{100, 83, 161}

\graphicspath{{figures/}} 
\begin{document}

\newcommand{\be}   {\begin{equation}}
\newcommand{\ee}   {\end{equation}}
\newcommand{\ba}   {\begin{eqnarray}}
\newcommand{\ea}   {\end{eqnarray}}
\newcommand{\ve}   {\varepsilon}
\newcommand{\Dis}  {\mbox{\scriptsize dis}}

\newcommand{\state} {\mbox{\scriptsize state}}
\newcommand{\band} {\mbox{\scriptsize band}}

\usetikzlibrary{arrows.meta, calc, positioning, shadows.blur}

\definecolor{hot1}{RGB}{255,37,56}   
\definecolor{hot2}{RGB}{140,16,16}   
\definecolor{cold1}{RGB}{4,18,189}   
\definecolor{cold2}{RGB}{6,44,112}   
\definecolor{axisgray}{RGB}{40,40,40} 

\tikzset{
  >=Latex,
  axis/.style={line width=1.2pt, draw=axisgray},
  cycle/.style={line width=1.1pt, draw=black},
  dashB/.style={draw=black!60, line width=0.9pt, dash pattern=on 2.2pt off 2.2pt},
  flowarrow/.style={-{Latex[length=4mm]}, line width=.11pt, draw=black},
  heatarrowH/.style={-{Latex[length=4mm]}, line width=2.5pt, draw=hot1},
  heatarrowC/.style={-{Latex[length=4mm]}, line width=2.5pt, draw=cold1},
  pt/.style={circle, inner sep=1.7pt, fill=black},
 resboxH/.style={
    rounded corners=2pt, 
    blur shadow={shadow blur steps=1, shadow xshift=-.1pt, shadow yshift=-0.pt, shadow opacity=0}, 
    inner sep=3.5pt,
    left color=hot2, right color=hot1,shadow scale=0.95, draw=white!70, line width=0.5pt
},
resboxC/.style={
    rounded corners=2pt, 
blur shadow={shadow blur steps=1, shadow xshift=-.1pt, shadow yshift=-0.pt,      shadow scale=0.95, shadow opacity=0.3},
    inner sep=3.5pt,
    left color=cold2, right color=cold1, draw=white!70, line width=0.5pt
}
}

\title{Reaching maximum efficiency in quantum Stirling engines using multilayer graphene}

\author{Bastian Castorene}
\email{bastian.castorene.c@mail.pucv.cl}
\affiliation{Instituto de Física, Pontificia Universidad Católica de Valparaíso, Casilla 4950, 2373223 Valparaíso,
Chile}
\affiliation{Departamento de Física, Universidad Técnica Federico Santa María, 2390123 Valparaíso, Chile}
\author{Francisco J. Peña} 
\affiliation{Departamento de Física, Universidad Técnica Federico Santa María, 2390123 Valparaíso, Chile}

\author{Eric Su\'arez Morell} 
\affiliation{Departamento de Física, Universidad Técnica Federico Santa María, 2390123 Valparaíso, Chile}

\author{Caio~Lewenkopf} 
\affiliation{Instituto de F\'isica, Universidade Federal do Rio de Janeiro, Rio de Janeiro, RJ 21941-972, Brazil}
 
\author{Martin HvE Groves} 
\affiliation{Instituto de Física, Pontificia Universidad Católica de Valparaíso, Casilla 4950, 2373223 Valparaíso,
Chile}
\affiliation{Departamento de Física, Universidad Técnica Federico Santa María, 2390123 Valparaíso, Chile}

\author{Natalia Cortés} 
\affiliation{Departamento de Física, Universidad Técnica Federico Santa María, 2390123 Valparaíso, Chile}
 
 \author{Patricio Vargas}
\affiliation{Departamento de Física, Universidad Técnica Federico Santa María, 2390123 Valparaíso, Chile}

\date{\today}

\begin{abstract}
In this work, quantum Stirling engines based on monolayer, AB-stacked bilayer, and ABC-stacked trilayer graphene under perpendicular magnetic fields are analyzed. Performance maps of the useful work \((\eta W)\) reveal a robust optimum at low magnetic fields and moderately low temperatures, with all stackings capable of reaching Carnot efficiency under suitable configurations. The AB bilayer achieves this across the broadest parameter window while sustaining finite work, the monolayer exhibits highly constrained regimes, and the trilayer shows smoother trends with sizable \(\eta W\). These results identify multilayer graphene, particularly the AB bilayer, as a promising platform for efficient Stirling engines, while also highlighting the versatility of the monolayer in realizing all four operational regimes of the Stirling cycle.
\end{abstract}
\maketitle
\section{Introduction}

Quantum thermodynamics examines how the fundamental principles of energy transfer, heat exchange, and entropy apply to small systems where quantum effects such as coherence, entanglement, spin correlations, and discrete energy levels give rise to thermodynamical properties with no classical counterpart, and where quantum statistics provides a natural framework to describe these phenomena~\cite{vinjanampathy2016quantum, kosloff2013quantum, deffner2019quantum,
millen2016perspective, binder2015quantum,
binder2018thermodynamics,
myers2022quantum,
Zanin2019,
Vieira2023,
WOS:000450309300001,
Castorene2,
WOS:000371827800005,
WOS:000351060000051}.

In this context, considerable attention has been devoted to understanding how quantum systems, such as trapped ions, superconducting qubits, quantum dots \cite{
esposito2010quantum, 
erdman2017thermoelectric, 
josefsson2018quantum, 
josefsson2019optimal, 
josefsson2020double,
du2020quantum}, 
and low-dimensional electronic materials and their diverse behaviors can be harnessed to implement heat engines capable of producing useful work. Among quantum thermal machines, the Stirling cycle has attracted particular interest due to its potential to achieve high efficiency in regimes where thermalization and isochoric processes can be controlled with precision~\cite{saygin2001quantum, 
sisman2001efficiency,
cruz2023quantum, 
raja2021finite, 
Cakmak2023,
castorene1,
WOS:000072570800004,
WOS:001439328500001,
WOS:000208227100001,
WOS:001414561800015,
araya,
WOS:000657177000003,
WOS:001316246400004}.
In light of these considerations, theoretical and experimental investigations have shown that quantum heat engines based on discrete energy spectra may surpass their classical counterparts in specific regimes, particularly when the working substance exhibits a tunable energy level structure and distinctive statistical properties~\cite{
pena2019magnetic,
deffner2018efficiency}.

Graphene and its multilayer stacks, when subjected to perpendicular magnetic fields 
\cite{castro2009electronic, McCann2006, Min2008, Yuan2011, Goerbig2011} stand out as promising platforms for quantum heat engines.
Beyond their ease of fabrication and high degree of experimental control~\cite{
koshino2008magneto,
nakamura2008electric,
WOS:001231812100001,
WOS:001197413400001,
WOS:000302524600034,
bao2011stacking}, 
manifestations of their quantized spectrum are robust against temperature, since  
these systems exhibit quantum Hall plateaus even at room temperatures under sufficiently high magnetic fields \cite{Novoselov2007roomT-QHE} -- a critical feature for practical applications. 
Furthermore, the large degeneracy of the Landau levels (LL) involves a macroscopic number of participating electrons, $N$, even in micron-size flakes \cite{Goerbig2011}
This provides a unique framework to explore quantum corrections to thermodynamic quantities stemming from the discrete Landau-level spectrum, while remaining within the thermodynamic limit.
Consequently, the small-$N$ thermal fluctuations typically found in few-body systems (both classical and quantum) are suppressed, making it possible to isolate genuine quantum effects.
Very importantly, when driven by a varying magnetic field $B$, these quantum systems - unlike most others - avoid the dissipation associated to Landau-Zener transitions \cite{Kayanuma1998LandauZener, Dai2025LandauZener}, as Landau levels do not cross with changes in $B$.
Finally, like other two-dimensional materials, the chemical potential (or doping) can be precisely tuned via electrostatic gating, offering a robust mechanism for controlling the engine's working fluid.

Several studies have proposed graphene-based systems, including quantum dots and multilayer structures, as promising candidates for quantum engines~\cite{
mani2019designing, 
singh2021magic, 
myers2023multilayer,
WOS:000964790300001,
WOS:000387051100002,
WOS:000354339000044,
WOS:000309268500009, 
cjlz-lrd6}. 
While these pioneering works offer valuable insights, they predominantly rely on Boltzmann statistics~\cite{pena2020quasistatic,myers2023multilayer,myers2022quantum,singh2021magic}. 
By neglecting the fermionic nature of the electronic working fluid and assuming charge neutrality, the lack of quantitative accuracy of their results and the narrow focus of their setting call for a reexamination.

This study examines a quantum Stirling cycle implemented in monolayer, AB-stacked bilayer, and ABC-stacked trilayer graphene under perpendicular magnetic fields as the working medium. 
Employing a low-energy Landau-level description and Fermi--Dirac statistics at finite doping, we correct the previous studies, expand they scope beyond the charge-neutrality point, and provide a thermodynamic analysis more closely aligned with experimentally accessible conditions \cite{PhysRevB.93.115424,rp3q-svws}. 
Within a grand-canonical framework, we evaluate key observables such as internal energy and entropy while explicitly accounting for the degeneracies specific to each multilayer configuration.

A central motivation of this work is to clarify how the microscopic Landau--level structure of mono-, bi-, and trilayer graphene shapes the performance of a quantum Stirling cycle. 
We show that the stacking-dependent spectra of these multilayers can naturally produce parameter regimes in which the Stirling efficiency approaches the Carnot bound while still delivering a finite amount of energy per cycle in the reversible, quasistatic limit (and thus zero power), without relying on quantum criticality or ideal regenerative mechanisms. 
This highlights the role of stacking order as a physically meaningful variable for controlling the thermodynamic response of quantum heat engines.

\begin{figure}[t]
    \centering
    \includegraphics[width=.99\columnwidth]{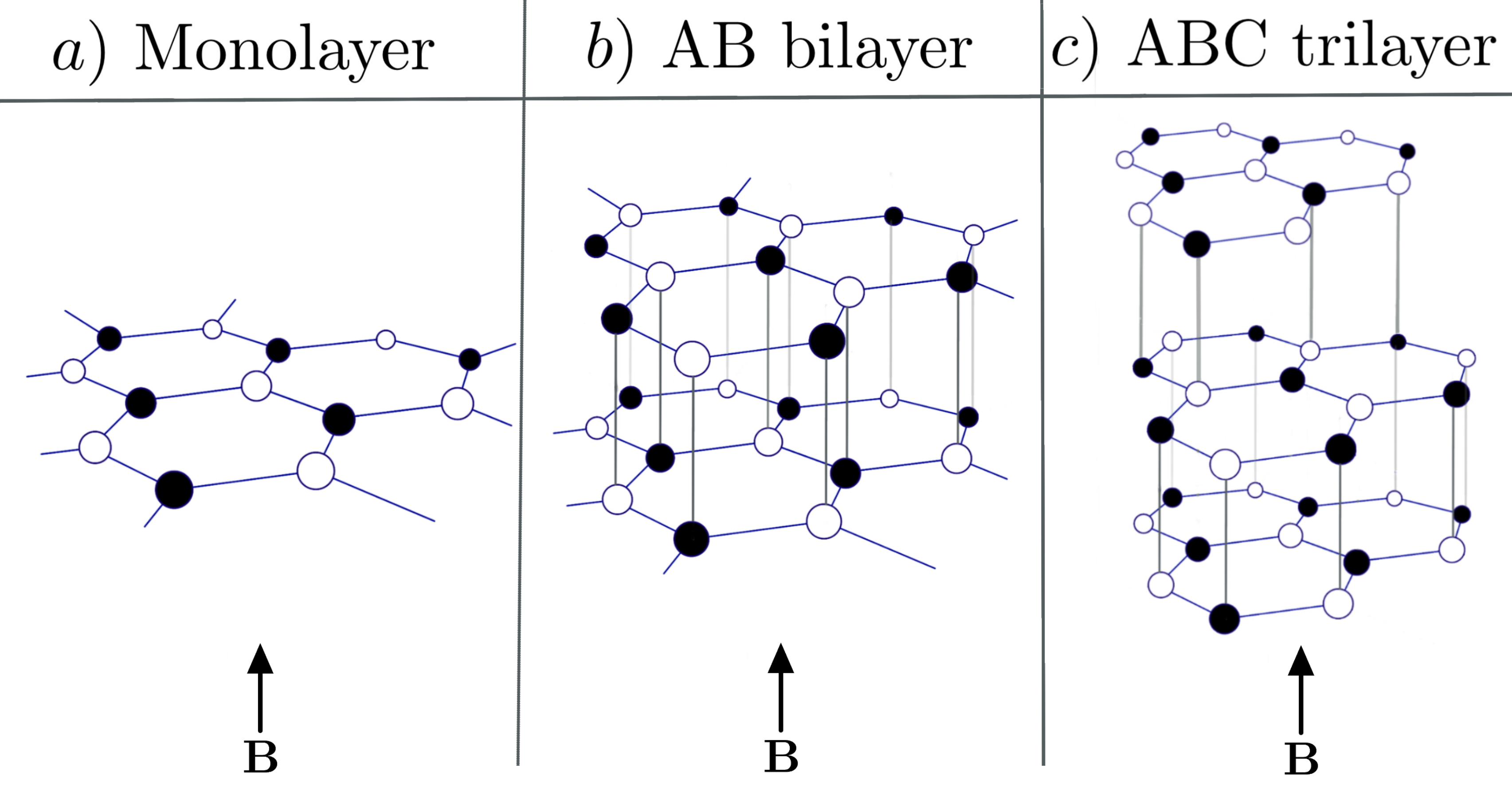}
    \caption{Schematic representation of the lattice structure of graphene multilayer systems (under a perpendicular homogeneous magnetic field \(B\)). Panel (a) shows the monolayer, panel (b) the AB-stacked bilayer, and panel (c) the ABC-stacked trilayer.}
    \label{fig:Diagrama_Grafenos}
\end{figure}

The structure of the paper is as follows. 
Section \ref{sec:method_and_methods} introduces the theoretical framework and the Landau level spectra for graphene with different multilayer stackings. 
Section \ref{Thermodynamic_Description} derives the thermodynamic functions required to implement the Stirling cycle. 
Section \ref{sec:results} presents the analysis of efficiency and work output across different regimes. 
Finally, Sec.~\ref{sec:conclusion} presents the conclusions with a summary and possible extensions.

\section{Theory and methods}
\label{sec:method_and_methods}

\subsection{Landau Level Spectrum in Multilayer Graphene}

At low electronic doping, the low-energy LL spectra of mono-, AB-bilayer, and ABC-trilayer graphene in a perpendicular magnetic field (as shown in Fig.~\ref{fig:Diagrama_Grafenos}) can be cast into a single compact form that highlights the multilayer scalings \cite{McCann2006,Min2008,Yuan2011,castro2009electronic,bao2011stacking}.

Introducing a stacking index \(J\) (\(J=1\) monolayer, \(J=2\) AB bilayer, \(J=3\) ABC trilayer), we write
\begin{equation}
\varepsilon^{(J)}_{n,\pm}
=
\pm \hbar \omega_c^{(J)}
\sqrt{\prod_{j=0}^{J-1}(n-j)},
\qquad
n=0,1,2,\ldots,
\label{energia_general}
\end{equation}
where $n$ is the LL index and $\pm$ indicate conduction and valence band LLs. 
The $J$-dependent cyclotron frequency \(\omega_c^{(J)}\) encodes the microscopic 
{parameters} 
of each system. 
{As we show below,} Eq.~\eqref{energia_general} yields the characteristic magnetic-field scaling
\begin{equation}
\varepsilon^{(J)}_{n} \propto B^{J/2},
\end{equation}
and predicts \(J\) zero-energy modes (\(n=0,1,\ldots,J-1\)). 
Consequently, the effect spin, valley and zero-energy level degeneracies results in a combined degeneracy \(g_n^{(J)}=4\) for all levels with $n\ge 1$ and 
    \begin{align}
        g_0^{(1)} = 4, \qquad g_0^{(2)} = 8, \qquad g_0^{(3)} = 12.
    \end{align}
for $n=0$.

For the three cases addressed here, the corresponding 
{LL energies are given by}
\begin{align}
J=1:\quad 
\varepsilon^{(1)}_{n,\pm} &= \pm \hbar \omega_c^{(1)}\sqrt{n},
\label{energia_mono}\\
\hbar\omega_c^{(1)} &= \sqrt{2 e \hbar v_F^2 B} {\,\approx 36.3\, \sqrt{B} \, [\mathrm{meV}],}
\nonumber\\[4pt]
J=2:\quad
\varepsilon^{(2)}_{n,\pm} &= \pm \hbar \omega_c^{(2)}\sqrt{n(n-1)},
\label{energia_bi}\\
\hbar \omega_c^{(2)} &= \frac{2e\hbar v_F^2 B}{t_\perp} {\,\approx 4.77 \, B\,[\mathrm{meV}],}
\nonumber\\[4pt]
J=3:\quad
\varepsilon^{(3)}_{n,\pm} &= \pm \hbar \omega_c^{(3)}\sqrt{n(n-1)(n-2)},
\label{energia_tri}\\
\hbar\omega_c^{(3)} &= 
\frac{\left(2 e B \hbar v_F^2\right)^{3/2}\sqrt{\hbar}}{t_\perp^2} {\,\approx 0.771 \, B^{3/2} \,[\mathrm{meV}],}
\nonumber
\end{align}
{
where the magnetic field $B$ is measured in tesla. 
Here, $v_F$ denotes the Fermi velocity of graphene and $t_\perp$ represents the interlayer hopping amplitude for each system, which characterizes the electronic coupling between adjacent graphene layers. 
Equations (4) to (6) establish distinct characteristic scales of the Landau-level spectra in multilayer graphene systems.
These differences are ultimately manifest in the thermodynamic properties discussed next.}

\section{Thermodynamic Description of Multilayer Graphene Systems Under Magnetic Field} \label{Thermodynamic_Description}
\subsection{Grand canonical formalism}

In recent studies using graphene as a thermal machine, the thermodynamic behavior of graphene multilayers has been investigated through thermal states based on the canonical ensemble and Landau-level spectra \cite{cjlz-lrd6,myers2023multilayer}, neglecting the fermionic nature of the working fluid.
In contrast, in this manuscript the thermodynamic properties of electrons in graphene multilayer systems under perpendicular magnetic fields are naturally described within the grand-canonical ensemble using Fermi–Dirac statistics, where both the temperature $T$ and the chemical potential $\mu$ serve as control parameters.

In this framework, the central quantities of interest are the internal energy \(U\), the entropy \(S\), and the particle number \(\langle N \rangle\), which together determine the efficiency and work output of thermal cycles.  

 \begin{figure*}[t]
  \centering
  \begin{subfigure}[b]{0.32\textwidth}
    \includegraphics[width=\textwidth]{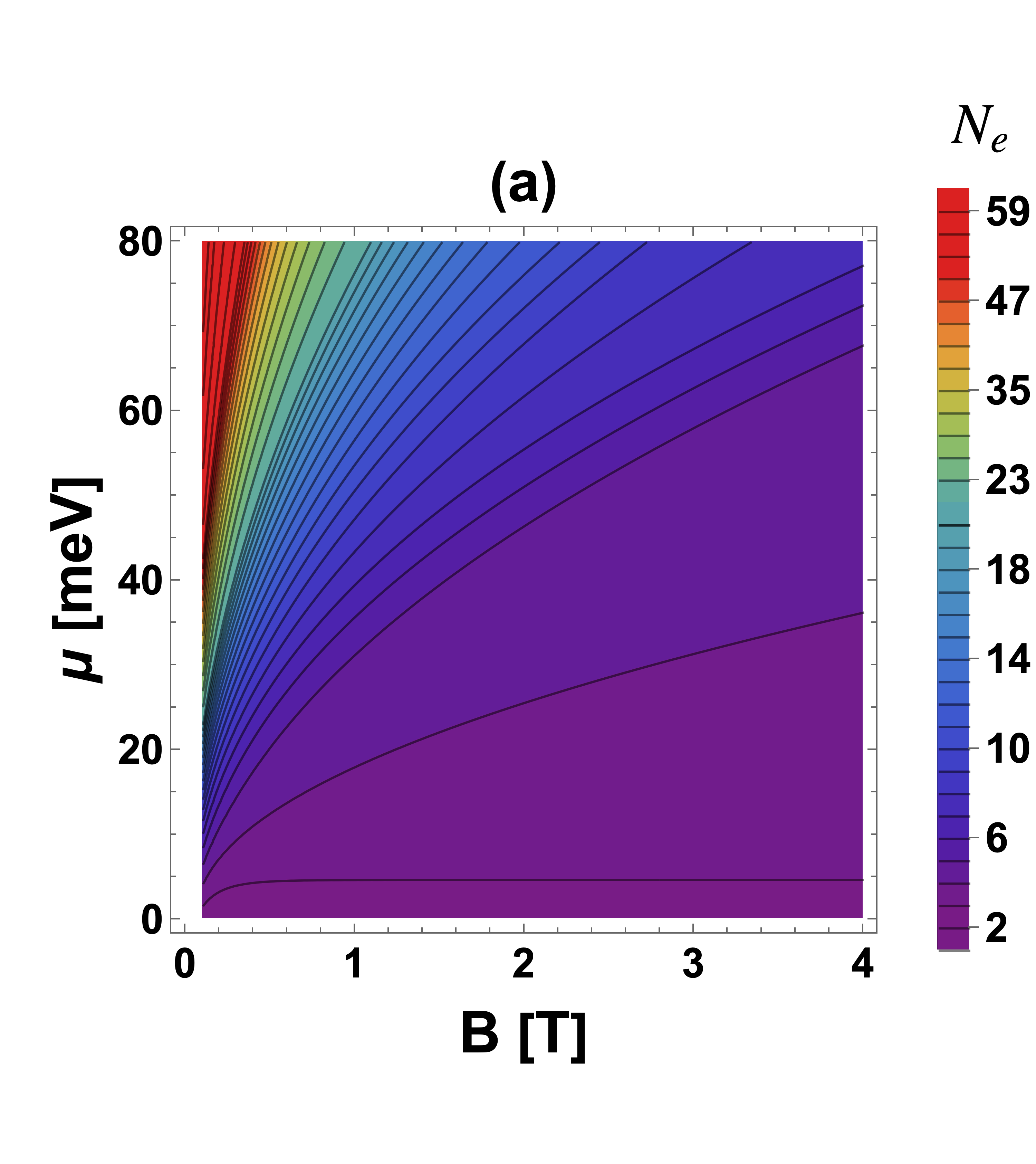}
  \end{subfigure}
  \hfill
  \begin{subfigure}[b]{0.32\textwidth}
    \includegraphics[width=\textwidth]{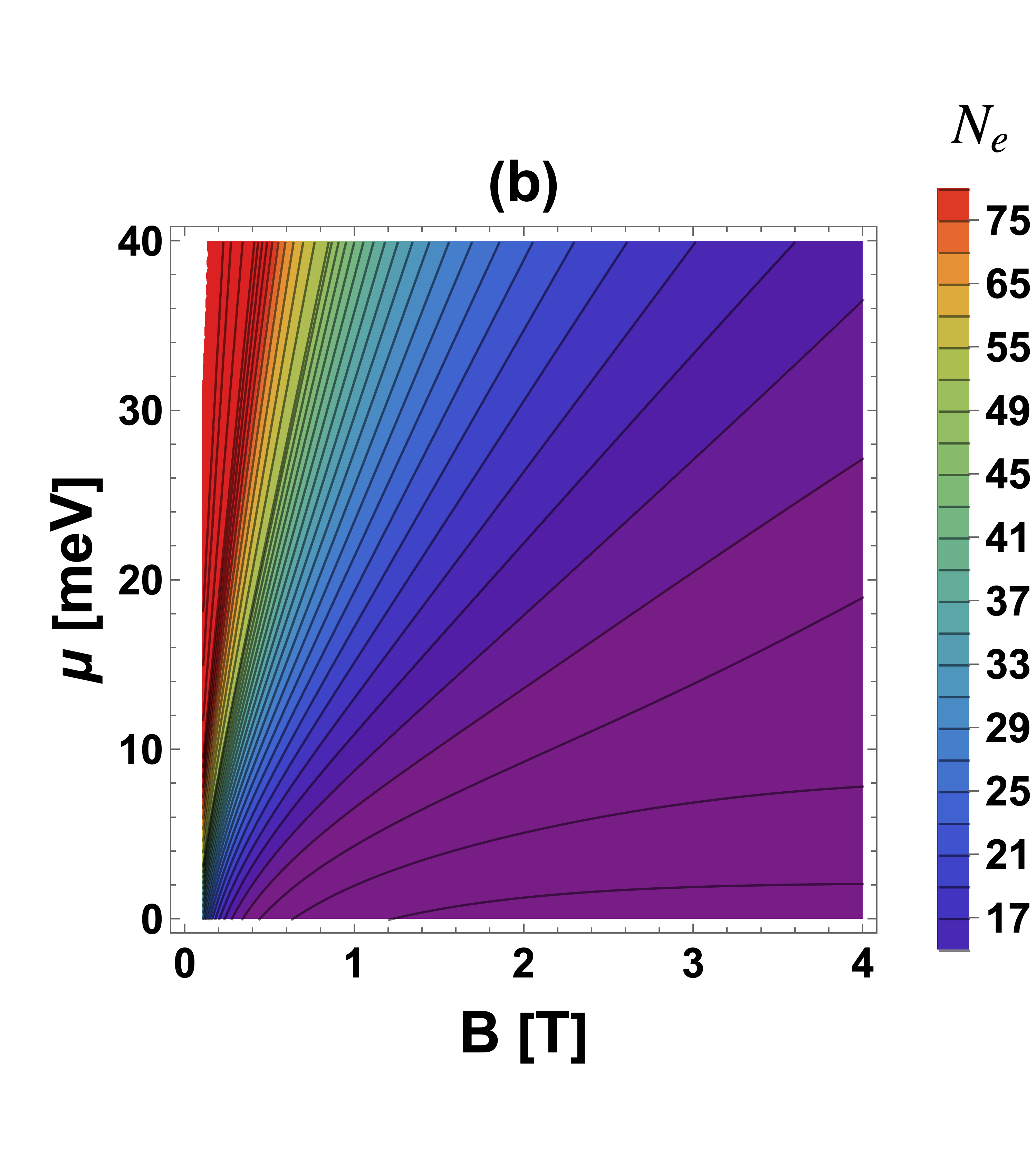}
  \end{subfigure}
  \hfill
  \begin{subfigure}[b]{0.32\textwidth}
    \includegraphics[width=\textwidth]{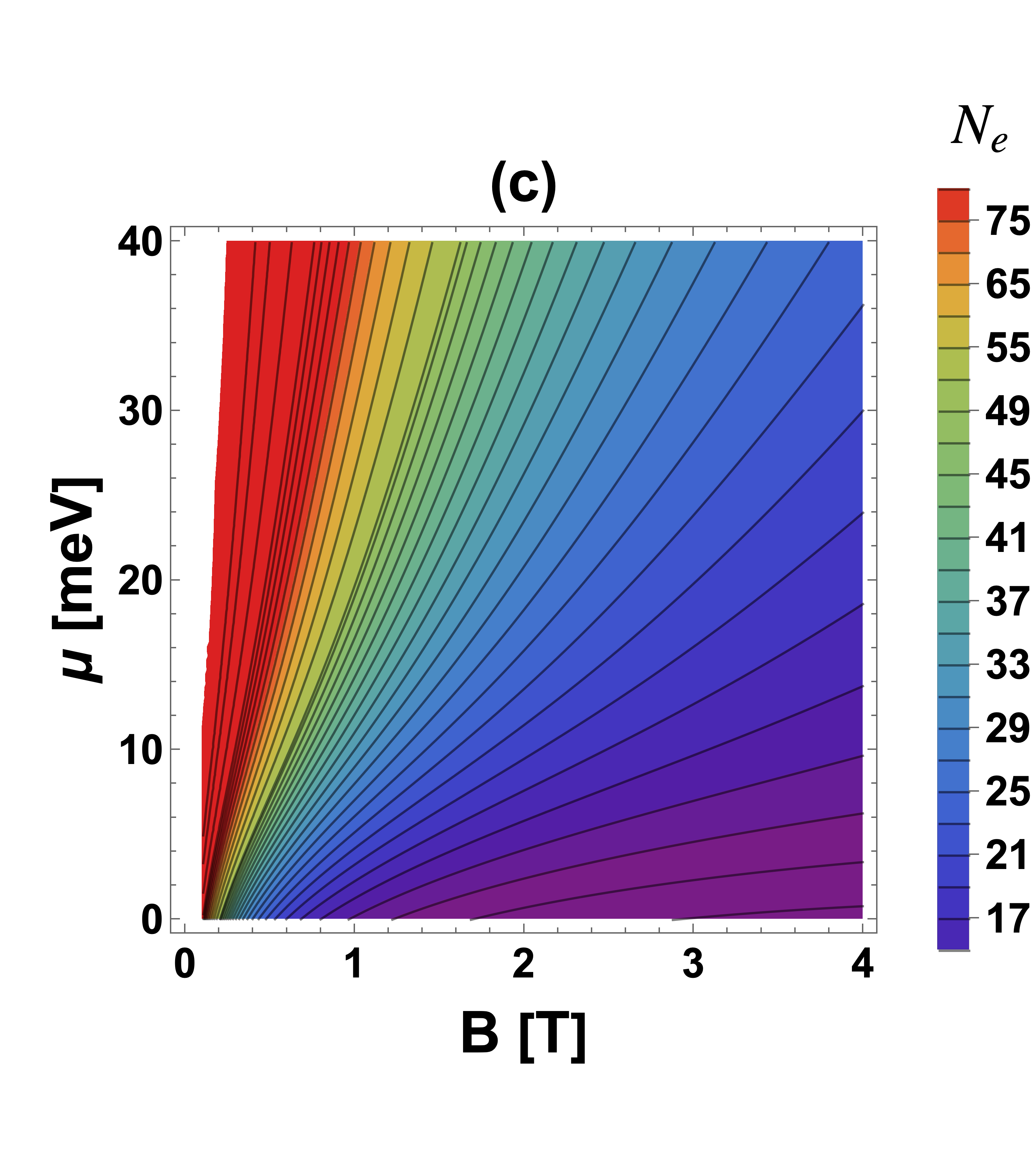}
  \end{subfigure}
  \caption{Color map of \(  N_e \) as a function of magnetic field \( B\, [T] \) and chemical potential \( \mu\ \) [meV] for (a) monolayer, (b) bilayer and (c) trilayered graphene system under a fixed temperature \( T =50  \) K.}
  \label{fig:3capas}
\end{figure*} 
For multilayer graphene with quantized Landau levels \(\varepsilon_{n,\pm}^{(J)}\),
the grand-canonical partition function can be written explicitly as
\begin{equation}
    \mathcal{Z}^{(J)} = \prod_{n,\pm} \left[ 1 + e^{-\beta \left( \varepsilon_{n,\pm}^{(J)} - \mu \right)} \right]^{g_{n}^{(J)}},
\end{equation}

Taking the degeneracies into account, the grand potential can be expressed as
\begin{equation}
    \Omega^{(J)} = -k_B T \sum_{n,\pm} g_{n}^{(J)} 
    \ln \left[ 1 + e^{-\beta \left( \varepsilon_{n,\pm}^{(J)} - \mu \right)} \right],
\end{equation}
where  $\beta = \frac{1}{k_B T}$.
The grand potential gives
\begin{align}
    \langle N \rangle = &- \frac{\partial \Omega^{(J)}}{\partial \mu} \eval{}_{T}, 
    \,\,
    S = - \frac{\partial \Omega^{(J)}}{\partial T}\eval{}_{\mu}, 
    \nonumber \\ 
    U = &\;\Omega^{(J)} + TS + \mu \langle N \rangle.
\end{align}

Equivalently, one may start from the Fermi–Dirac distribution, which gives the occupation of each single-particle state $f(\varepsilon,\mu,T) = [e^{\beta (\varepsilon - \mu)} + 1]^{-1}$    
together with the density of states (DOS),
\begin{equation}
    \nu^{(J)}(\varepsilon) = \sum_{n,\pm} g_{n}^{(J)} \, \delta(\varepsilon - \varepsilon_{n,\pm}^{(J)}),
\end{equation}
which does not yet include the LL orbital degeneracy, that is to be discussed in the next section. 
These allow the same observables to be written as
\begin{align}
    \expval{ N (T,B, \mu)} &= \int \limits_{-\infty}^{\infty} d\varepsilon\, \nu^{(J)}(\varepsilon)\, f(\varepsilon,\mu,T), \\
    U(T,B, \mu) &= \int\limits_{-\infty}^{\infty} d\varepsilon\, \nu^{(J)}(\varepsilon)\, \varepsilon\, f(\varepsilon,\mu,T), \\
    S(T,B, \mu) &= -k_B \!\!\int \limits_{-\infty}^{\infty} \!d\varepsilon\, \nu^{(J)}(\varepsilon)\,
    [ f \ln f + 
    \nonumber\\ & \;\;\;\;\;\;\;\;\;\;\;\;\;\;\;\;\;\;\;\;\;\;\;\;\;
    (1-f) \ln (1-f) ].
\end{align}

Both approaches are formally equivalent: the discrete-level formulation based on the grand potential and the continuum representation using the DOS combined with the Fermi–Dirac distribution, both yielding to identical thermodynamic quantities. 
It is worth emphasizing that the DOS representation makes explicit the role of Landau-level degeneracies, particularly the zero-energy states in all systems.  
We adopt the latter approach for our calculations. 
To guarantee numerical stability across the entire parameter space, the sums over Landau levels are truncated at a cutoff index of $n=2000$, or when the relative difference between consecutive evaluations falls below $10^{-6}$. 
Crucially, for the specific temperature and magnetic field regimes considered, thermal population is confined to the lowest levels, with contributions from $n>20$ being exponentially suppressed. 
Consequently, the effective number of participating states is limited to $n \lesssim 20$, which is fully consistent with the $\delta$-function DOS representation employed in this work \cite{Vidarte2022}.

\subsection{Energy Scales and Parametrization of the Chemical Potential}
Having established the theoretical framework, we now estimate the order of magnitude of the energy separation between the ground state and the first accessible Landau excitation as the magnetic field \(B\) is varied. This quantity {$\hbar\omega_c^{(J)}$}
sets the relevant energy scale for the thermodynamic analysis, since it determines the temperatures required to thermally populate excited states as the field is varied \cite{castro2009electronic,McCann2006,bao2011stacking,Min2008}.

For reference to experimental scales, we provide an estimate obtained from the magnetic length, \(\ell_B = \sqrt{\hbar/(eB)}\) \cite{castro2009electronic, Goerbig2011}. By defining \(N_e\) as the number of electrons within the magnetic length area \(\pi \ell_B^2\), the sheet density can be written as
\begin{align}
    n_{2\mathrm D} \approx \frac{eB}{\pi\hbar}\,N_e \simeq 4.86\times 10^{10}\,N_e\,B~\mathrm{cm^{-2}}.
\end{align}

Typical electron concentrations in graphene vary according to the substrate and doping, with values ranging from near-neutral (around \(10^{12}\) cm$^{-2}$) up to \(10^{13}\) -- \(10^{14}\) cm$^{-2}$ for doped samples \cite{Novoselov2005, Liu2013}.
Consequently, a standard 1$\mu$m$^2$ sample with $n_{2\mathrm D} \approx 10^{13}$ cm$^{-2}$ contains $\simeq 10^5$ electrons in the conduction band.
For the purpose of our numerical analysis under modest magnetic fields ($B \approx 1$ T), these experimental scales constrain the modeled particle count to $N_e \lesssim 25$. 

Fig.~\ref{fig:3capas} provides an overview of the relation between the
magnetic field $B$, the chemical potential $\mu$, and the resulting particle
number $N_e$ for mono-, bi-, and trilayer graphene at fixed
temperature of $T= 50$ K . In this figure, $N_e$ is not fixed; instead, we map the full surface
$ N_e(50 ,B ,\mu ) $ in order to illustrate how the distinct Landau–level
structures of each multilayer system shape the dependence of the particle
number on $\mu$ and $B$. This exploration helps visualize the scaling behaviors
originating from the linear, quadratic, and cubic dispersions of the three
stackings.

To make a consistent comparison across all layers, the chemical potential is
determined self-consistently from the particle-number constraint
\begin{equation}
    N_e = \frac{\left\langle N(T,B,\mu(T,B)) \right\rangle}{\pi \ell_B^2} = \text{constant},
    \label{eq:N_e}
\end{equation}
so that $\mu(T,B)$ is obtained at every point of the thermodynamic cycle.
Within this framework, the fixed particle number $N_e$ fully determines the
temperature and magnetic-field dependence of the chemical potential, and no
additional analytical assumption regarding its $T$-dependence is imposed.

We emphasize that, for fixed $B$, the chemical potential also acquires a
(comparatively weaker) dependence on temperature, i.e., $\mu=\mu(T,B)$ is
determined self-consistently throughout the cycle. 
This temperature dependence is fully included in our numerical evaluation of the thermodynamic quantities.

\subsection{Thermodynamic Properties: Internal Energy and Entropy of Multilayer Graphene}\label{Thermodynamic-Properties}

\begin{figure}[h!]
    \centering
    \begin{subfigure}[t]{0.43\textwidth}
        \centering
        \includegraphics[width=\textwidth]{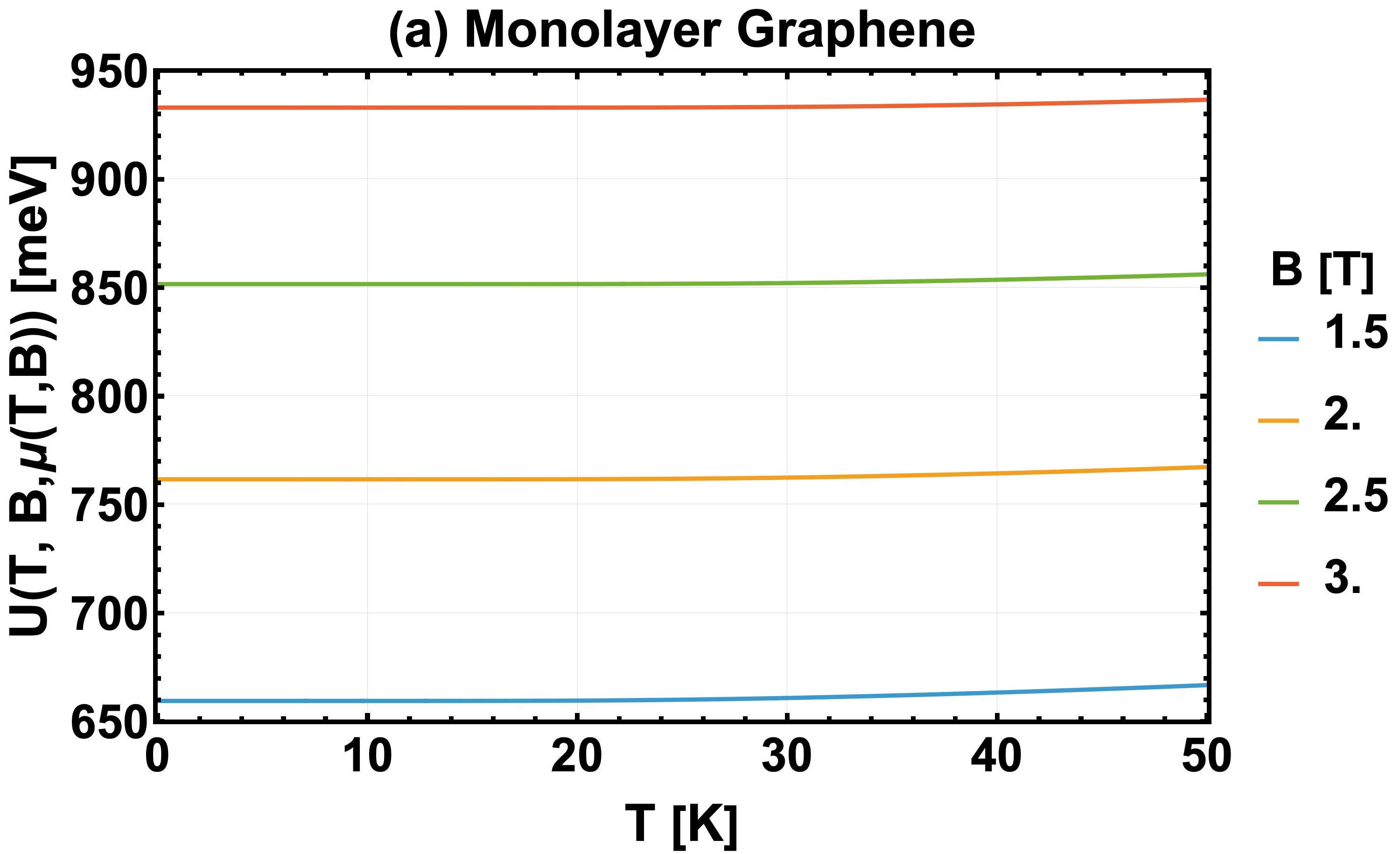}
    \end{subfigure}
    \hfill
    \begin{subfigure}[t]{0.43\textwidth}
        \centering
        \includegraphics[width=\textwidth]{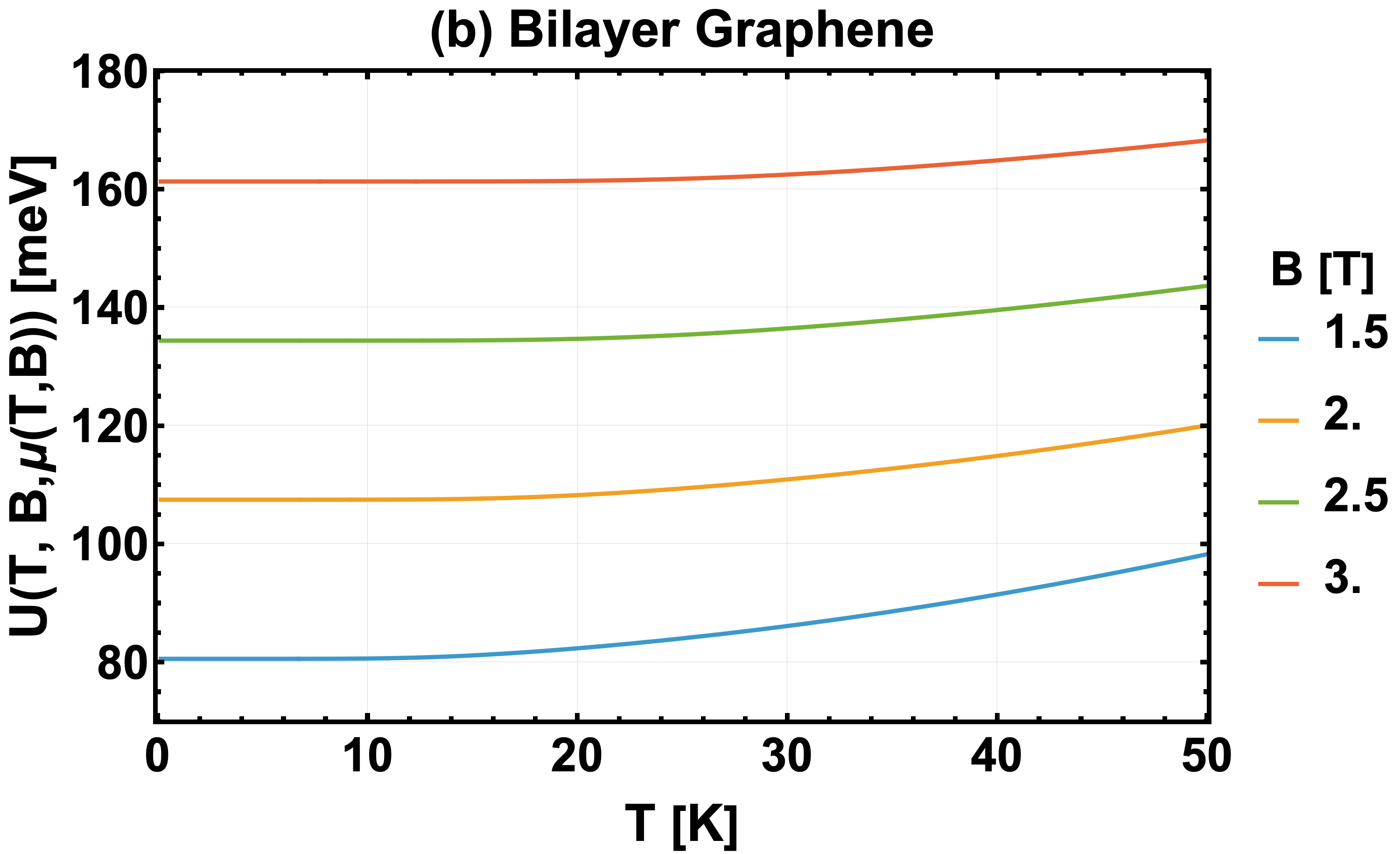}
    \end{subfigure}

    \begin{subfigure}[t]{0.43\textwidth}
        \centering
        \includegraphics[width=\textwidth]{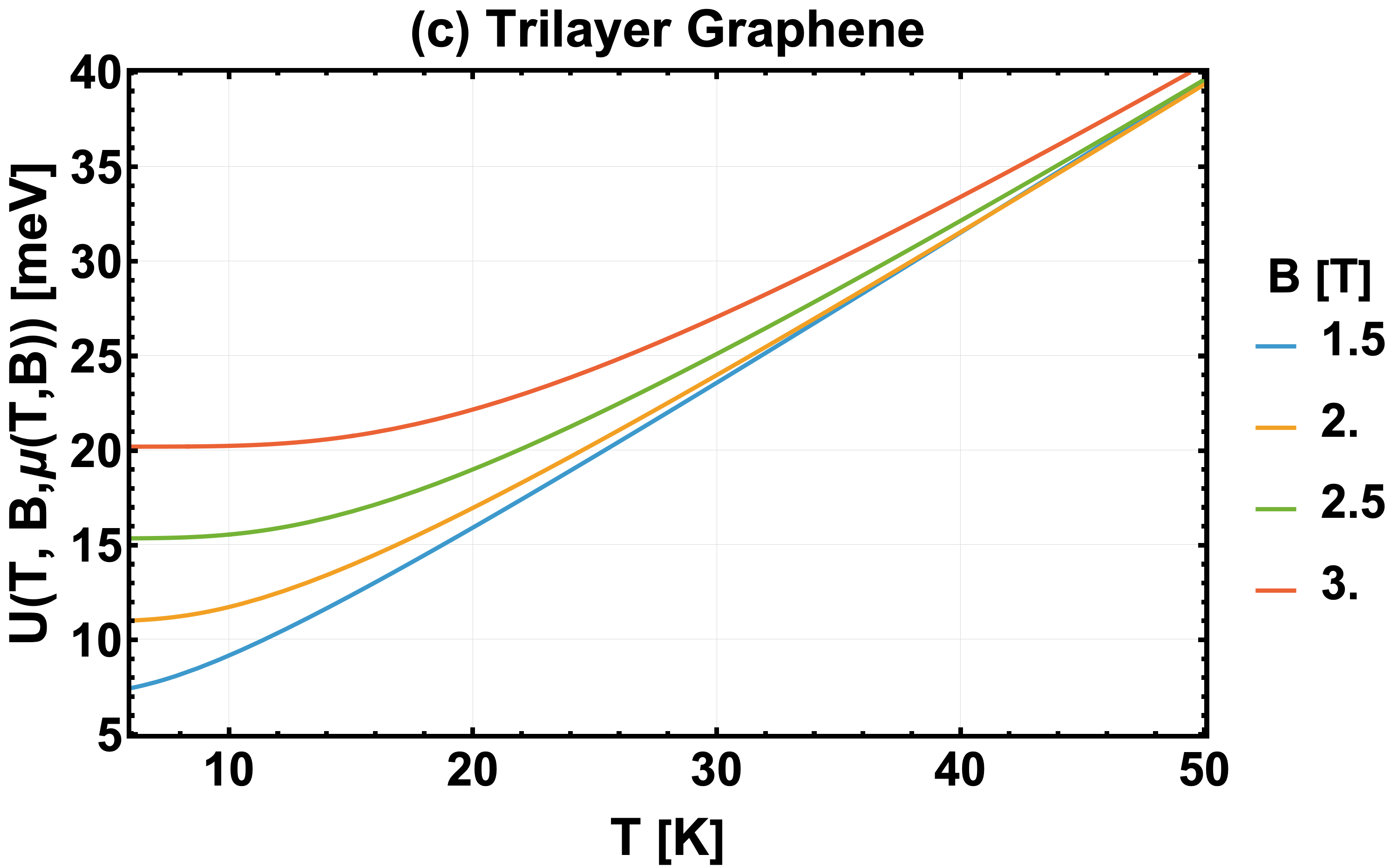}
    \end{subfigure}
    \caption{Internal energy \( U(T,B,\mu (T,B)) \) as a function of temperature for (a) monolayer, (b) bilayer, and (c) trilayer graphene at different magnetic fields, with the particle number fixed at \( N_e = 15 \).}
    \label{fig:internal_energyvsT}
\end{figure}

Having established the Landau-level spectra in Eqs.~\ref{energia_mono}, \ref{energia_bi}, and \ref{energia_tri}, together with the parametrized chemical potential of Eq.~\ref{eq:N_e} that ensures a fixed particle number across all three systems. 
We now turn to the analysis of thermodynamic quantities directly relevant for the operation of quantum heat engines. 
In particular, we investigate how each multilayer configuration responds under thermal cycling, an aspect of central importance for the implementation of the Stirling cycle. 

The internal energy \(U(T,B,\mu (T,B))\) of monolayer, bilayer, and trilayer graphene is presented in Fig.~\ref{fig:internal_energyvsT} as a function of temperature for several fixed magnetic-field values, at constant particle number \(N_e = 15\).

\begin{figure}[H]
    \centering
    \begin{subfigure}[t]{0.43\textwidth}
        \centering
        \includegraphics[width=\textwidth]{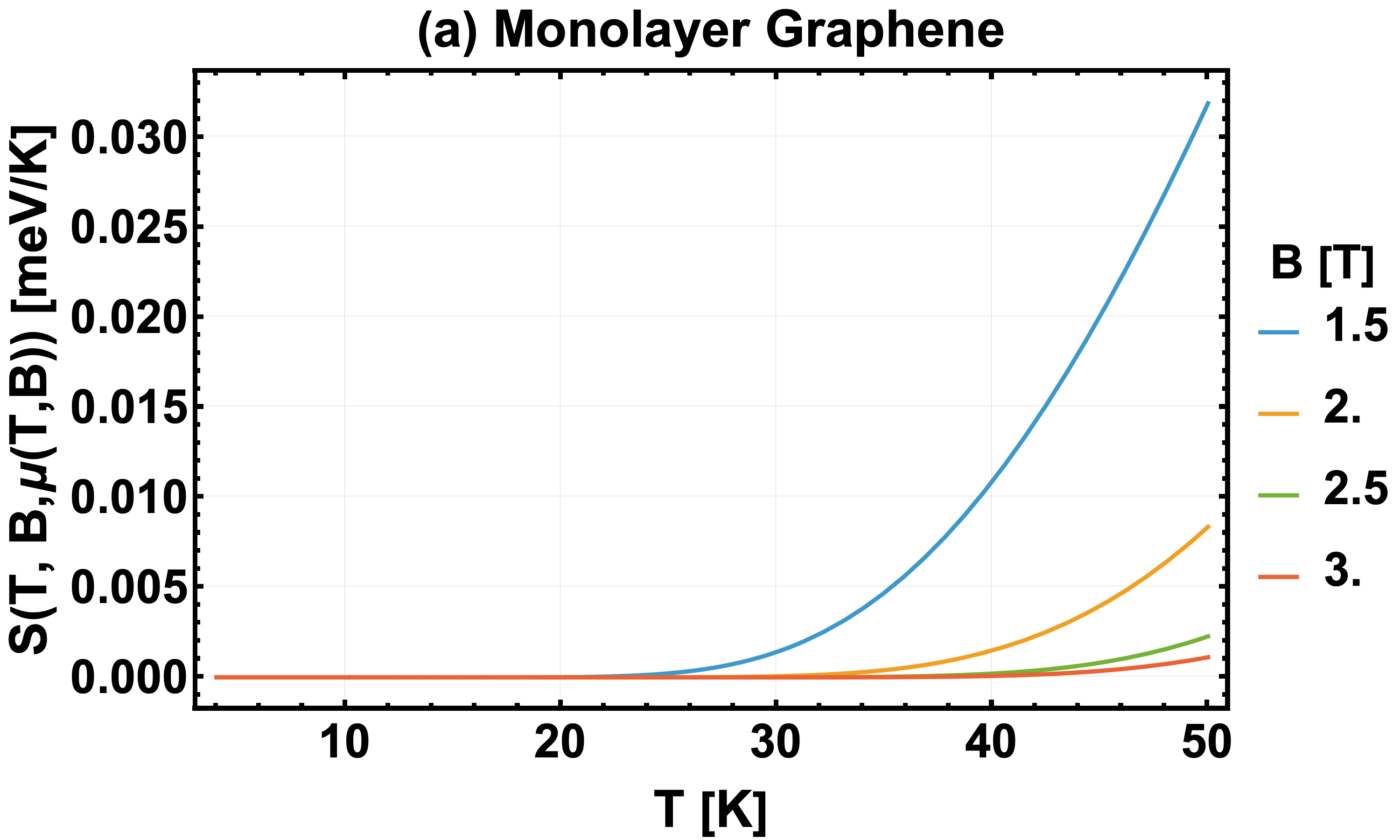}
    \end{subfigure}
    \hfill
    \begin{subfigure}[t]{0.43\textwidth}
        \centering
        \includegraphics[width=\textwidth]{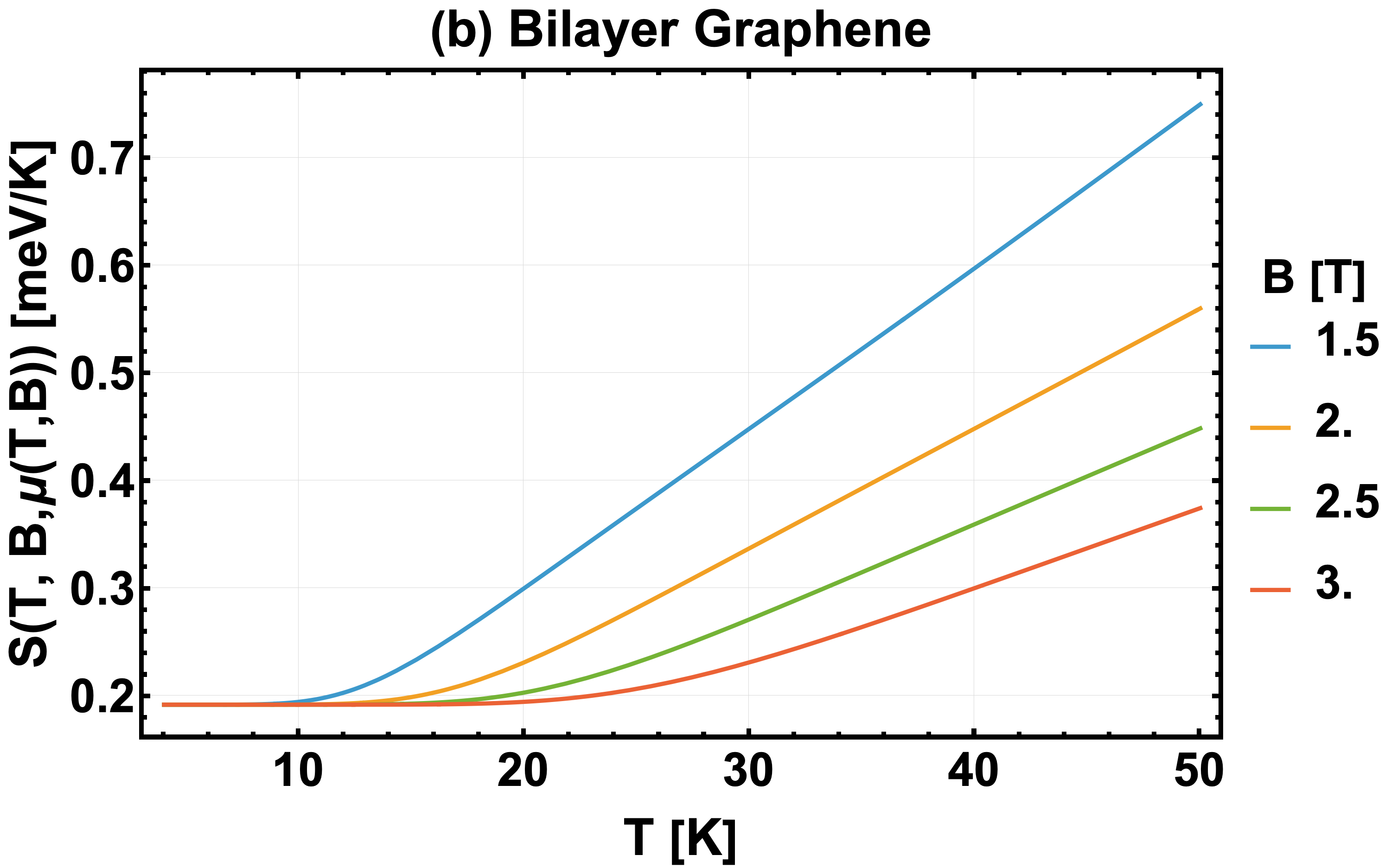}
    \end{subfigure}

    \vspace{0.5cm}

    \begin{subfigure}[t]{0.43\textwidth}
        \centering
        \includegraphics[width=\textwidth]{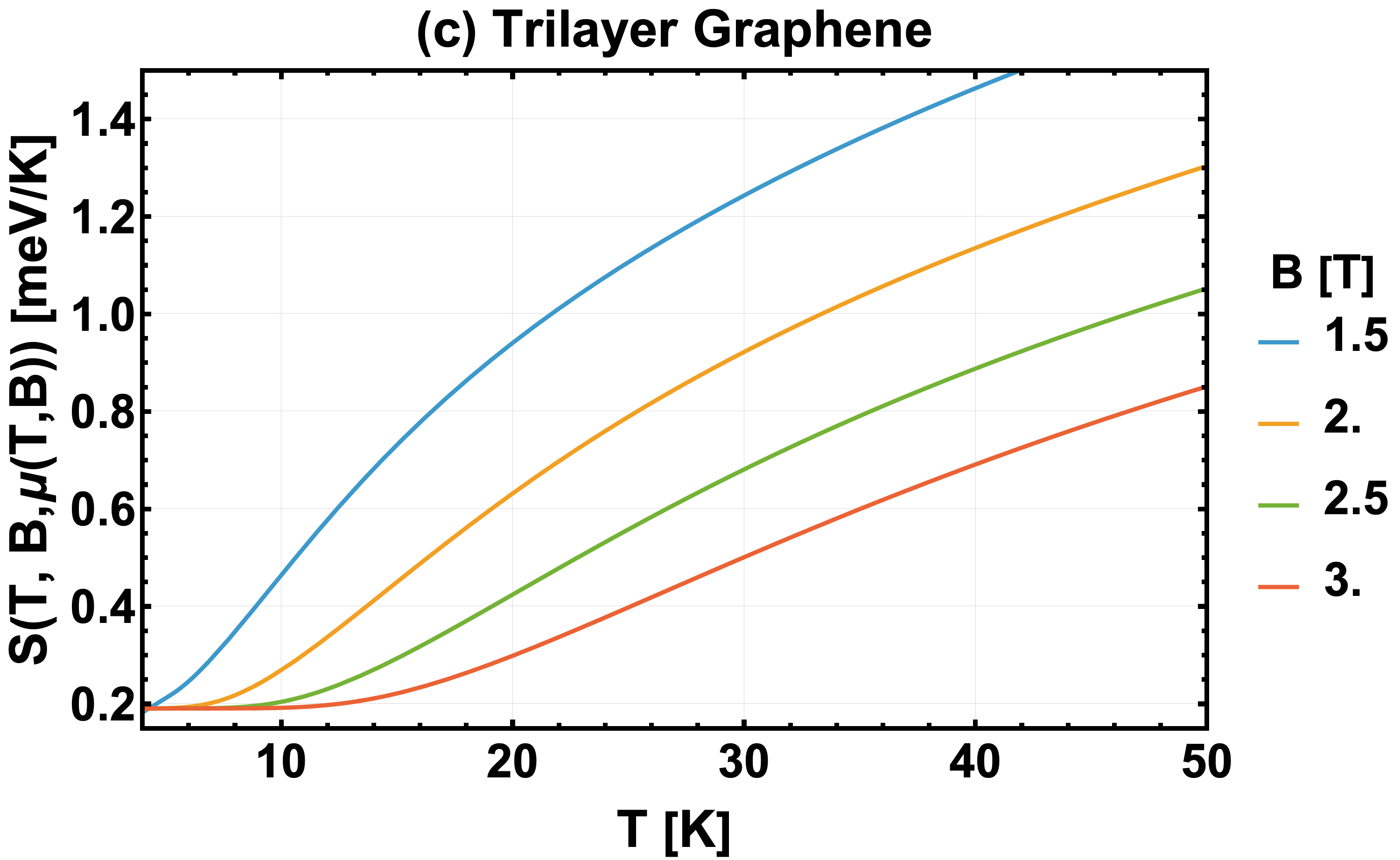}
    \end{subfigure}
\caption{Entropy \( S(T,B,\mu (T,B)) \) as a function of temperature for (a) monolayer, (b) bilayer, and (c) trilayer graphene at different magnetic fields, with the particle number fixed at \( N_e = 15 \).}
    \label{fig:entropy}
\end{figure}

In the monolayer case, shown in Fig.~\ref{fig:internal_energyvsT}(a), the internal energy increases moderately with temperature, following a smooth and weakly nonlinear trend. This behavior reflects the relativistic nature of Dirac fermions and the square-root spacing of the Landau levels. Increasing the magnetic field enhances the Landau-level separation, which produces an upward shift in the baseline of internal energy, while the overall temperature dependence remains predominantly sublinear. For the range of fields considered, the internal energy varies between approximately 600 and 950$\mathrm{meV}$. In Fig.~\ref{fig:internal_energyvsT}(b), the bilayer system exhibits a more pronounced curvature compared to the monolayer, particularly at intermediate temperatures. At lower magnetic fields, the internal energy rises more rapidly with temperature, consistent with the parabolic band structure and the enhanced density of states near charge neutrality. Because of the smaller Landau-level separations, the states are more easily populated than in the monolayer, which increases the sensitivity of the internal energy to thermal excitations. The overall magnitude of internal energy varies between approximately 80 and 170 [$\mathrm{meV}$] across the magnetic-field range considered.

In Fig.~\ref{fig:internal_energyvsT}(c), trilayer graphene exhibits a qualitatively distinct behavior. At high temperatures, all curves display an approximately linear increase in internal energy with comparable slopes, largely independent of the magnetic field. Similar to the monolayer and bilayer cases, the separation between different B values persists throughout the entire range, with the distinction being most evident at low temperatures, indicating that magnetic-field effects remain relevant despite thermal excitations. This behavior reflects the denser and more intricate low-energy spectrum of the trilayer, which enhances the internal energy under stronger fields. Overall, the temperature dependence of the internal energy in trilayer graphene arises from a competition between two effects: the magnetic field increases the spacing between Landau levels, thereby reducing the number of thermally accessible states and suppressing the growth of internal energy; in contrast, raising the temperature promotes the occupation of higher-energy states, leading to an overall increase in internal energy. Within the explored temperature window, no crossover or inversion is observed among the curves for different magnetic fields, suggesting that neither mechanism fully dominates.

The next discussion focuses on the entropy to further characterize the thermodynamic response. Fig.~\ref{fig:entropy} presents the temperature dependence of the entropy $S(T,B,\mu (T,B))$ for monolayer, bilayer, and trilayer graphene at different magnetic fields, with the particle number fixed at $N_e = 15$. While all three systems exhibit a monotonic increase of entropy with temperature, notable differences arise in magnitude, curvature, and sensitivity to the magnetic field, underscoring the distinct thermodynamic responses of each multilayer configuration.
In Fig.~\ref{fig:entropy}(a), the entropy of monolayer graphene remains negligible at low temperatures and begins to rise appreciably only above $T \simeq 25\,\mathrm{K}$, where thermal excitations populate higher Landau levels. The increase is markedly nonlinear, particularly at lower magnetic fields. This behavior reflects the relativistic Landau-level structure  where stronger magnetic fields enlarge the spacing between energy levels and thereby suppress thermal activation. Consequently, higher fields substantially reduce the entropy and produce a clear separation among the curves.

In Fig.~\ref{fig:entropy}(b), the bilayer case exhibits a more regular and approximately linear increase of entropy with temperature throughout the explored range. The entropy rises from about $0.2\,\mathrm{meV/K}$ to nearly $0.75\,\mathrm{meV/K}$ at the lowest field $B = 1.5\,\mathrm{T}$. The larger values compared to the monolayer reflect the denser Landau-level structure associated with the parabolic dispersion. Sensitivity to the magnetic field is also more pronounced, as stronger fields systematically shift the curves downward.

For the trilayer configuration shown in Fig.~\ref{fig:entropy}(c), the entropy starts near \(0.2\,\mathrm{meV/K}\) and grows beyond \(1.3\,\mathrm{meV/K}\) for \(B = 1.5\,\mathrm{T}\), displaying a rapid and nonlinear increase with temperature. The entropy curves exhibit a strong dependence on the magnetic field. This behavior originates from the cubic low-energy dispersion, which generates a high density of states and reduces the spacing between adjacent energy levels, effectively producing a quasi-continuous spectrum that remains thermally active even under stronger magnetic fields.

\subsubsection{Quantum Stirling Cycle}

\begin{figure}
    \centering
\resizebox{1.\linewidth}{!}{     \begin{tikzpicture}[x=0.65cm,y=0.65cm]
\draw[axis,->] (1,0) -- (9,0) node[ right=-3pt] { $ B$};
\draw[axis,->] (1,0) -- (1,6.3) node[above =-2.pt] { $ {S(T,B, \mu (T,B))}$};

\def\BL{3.0}
\def\BH{7.8}

\draw[dashB] (\BL,0) -- (\BL,6);
\draw[dashB] (\BH,0) -- (\BH,6);

\coordinate (A) at (\BH,3.6);
\coordinate (B) at (\BL,5.0);
\coordinate (C) at (\BL,2.3);
\coordinate (D) at (\BH,0.9);

\draw[cycle] 
  (A) .. controls ($(A)+(0.1,-0.)$) and ($(B)+(3.5,.5)$) .. (B);
\draw[cycle] (B) -- (C);
\draw[cycle] 
  (C) .. controls ($(C)+(3.5,-0)$) and ($(D)+(.1,.)$) .. (D);
\draw[cycle] (D) -- (A);

\draw[flowarrow] ($(A)!0.85!(B)$) ++(3.4,-0.53) -- ++(-0.3,0.2) ;
\draw[flowarrow] ($(B)!0.55!(C)$) -- ++(0,-.6);
\draw[flowarrow] ($(C)!0.6!(D)$) ++(-.6,0.58) -- ++(0.3,-.03) ;
\draw[flowarrow] ($(D)!0.45!(A)$) -- ++(0,1.0);

\node[ right=0pt of A] { $A$};
\node[ left=0pt of B] {$B$};
\node[ left=1pt of C] {$C$};
\node[ right=1pt of D] { $D$};

\node[below] at (\BL,-0.05) { $B_{L}$};
\node[below] at (\BH,-0.05) { $B_{H}$};

\node[resboxH] (TH) at (4.45,5.1) {\large $\textcolor{white}{T_{H}}$};
\node[resboxC] (TL) at (6.8,1.4) {\large $\textcolor{white}{T_{L}}$};
\draw[heatarrowH] (6.5, 5.1) -- ++(-1.1,-.7) node[pos=-1, below left=-1pt] { $Q_{AB}$};
\draw[heatarrowC] ($(B)!0.55!(C)$) ++(0.6,0.35) -- ++(-1.4,0) 
  node[pos=1.1,  above left=-10pt] { $Q_{BC}$};
\draw[heatarrowC] ($(C)!0.48!(D)$) ++(-.3,.8) -- ++(-1.1,-0.7)
  node[pos=1, below =-6pt] { $Q_{CD}$};
\draw[heatarrowH] ($(D)!0.87!(A)$) ++(0.66,-1.1) -- ++(-1.4,0) 
  node[pos=-.19, above right=-12pt] { $Q_{DA}$};

\foreach \P in {A,B,C,D} \fill ( \P ) node[pt] {};

\end{tikzpicture}}
    \caption{Diagram of the quantum Stirling cycle in terms of entropy $S(T,B,\mu (T,B))$ for fixed particle-number $N_e$ and the perpendicular magnetic field $B$. 
    Curves AB and CD correspond to the isotherms at $T = T_H$ and $T = T_L$, respectively.
    As the magnetic field increases at fixed temperature, the entropy decreases, and vice versa.}
    \label{fig:stirling_engine}
\end{figure}

The quantum Stirling cycle, illustrated in Fig.~\ref{fig:stirling_engine}, consists of four strokes: two isothermal and two isomagnetic processes. 
In this framework, the perpendicular magnetic field \(B\) acts as the external control parameter, while the working substance remains in thermal equilibrium with the reservoirs throughout the entire cycle. 
For each pair of values \((T,B)\), the chemical potential \(\mu(T,B)\) is determined self-consistently to keep the particle number \(N_e\) fixed, and the resulting value is then used in all thermodynamic quantities. 
The sequence of strokes is as follows:
\begin{itemize}
    \item \textbf{(A $\rightarrow$ B)} Isothermal expansion at temperature \(T_H\), with the magnetic field decreasing from \(B_H\) to \(B_L\),
    \item \textbf{(B $\rightarrow$ C)} Isomagnetic cooling at fixed \(B = B_L\), with the temperature lowered from \(T_H\) to \(T_L\),
    \item \textbf{(C $\rightarrow$ D)} Isothermal compression at temperature \(T_L\), with the magnetic field increasing from \(B_L\) to \(B_H\),
    \item \textbf{(D $\rightarrow$ A)} Isomagnetic heating at fixed \(B = B_H\), with the temperature raised from \(T_L\) back to \(T_H\).
\end{itemize}

The heat exchanged during each stage is expressed as
\begin{align}
    Q_{AB} &= T_H \big[ S( T_H,B_L,\mu( T_H,B_L)) - S( T_H,B_H,\mu ( T_H,B_H)) \big], \\
    Q_{CD} &= T_L \big[ S( T_L,B_H,\mu ( T_L,B_H)) - S( T_L,B_L,\mu ( T_L,B_L)) \big], \\
    Q_{BC} &= U( T_L,B_L,\mu ( T_L,B_L)) - U( T_H,B_L, \mu (T_H,B_L)), \\
    Q_{DA} &= U( T_H,B_H, \mu ( T_H,B_H)) - U( T_L,B_H, \mu ( T_L,B_H)),
\end{align}
where the labels \(L\) and \(H\) are used to distinguish the low and high values of 
$B$ and $T$. Invoking the first law of thermodynamics, the total heat absorbed $Q_{\rm in}$, the heat released $Q_{\rm out}$, and the net work $W$ per cycle can be expressed as
\begin{align}
    Q_{\text{in}} &= Q_{AB} + Q_{DA}, \\
    Q_{\text{out}} &= Q_{BC} + Q_{CD}, \\
    W &= Q_{\text{in}} + Q_{\text{out}}.
\end{align}

The convention adopted here is that \(Q > 0\) corresponds to heat absorbed by the system, while \(W > 0\) indicates net work performed by the system. Within this framework, the efficiency of the cycle is given by
\begin{equation}
    \eta = \frac{W}{Q_{\text{in}}} = 1 - \left| \frac{Q_{BC} + Q_{CD}}{Q_{AB} + Q_{DA}} \right|. \label{efi_eq}
\end{equation}

The physical significance of these expressions is clarified when related to the entropy behavior discussed in Sec.~\ref{Thermodynamic-Properties}, which shows that entropy increases as the external magnetic field decreases at fixed temperature. This guarantees that the isothermal expansion stroke (A~$\rightarrow$~B), performed from \( B_H \) to \( B_L \), corresponds to a positive entropy change \( \Delta S > 0 \), and thus a positive heat input \( Q_{AB} > 0 \). In contrast, the isothermal compression stroke (C~$\rightarrow$~D), carried out at lower temperature and in the reverse direction \( B_L \to B_H \), yields a negative entropy change and therefore a negative heat exchange \( Q_{CD} < 0 \). Consequently, the different operational regimes of the engine are governed entirely by the heat contributions during the isomagnetic strokes (B~$\rightarrow$~C and D~$\rightarrow$~A).
\begin{table}[t]
  \caption{Sign convention for heat and work: \( W>0 \) for work output, \( W<0 \) for input; \( Q>0 \) for heat absorbed, \( Q<0 \) for heat released by the system.}
  \label{tab:signs_work_heat}
  \begin{ruledtabular}
    \begin{tabular}{l|c|c|c}
      Operational regime & \(W\) & \(Q_{\mathrm{out}}\) & \(Q_{\mathrm{in}}\) \\
      \hline
      Engine       & \(>\!0\) & \(<\!0\) & \(>\!0\) \\
      Refrigerator & \(<\!0\) & \(>\!0\) & \(<\!0\) \\
      Accelerator  & \(<\!0\) & \(<\!0\) & \(>\!0\) \\
      Heater       & \(<\!0\) & \(<\!0\) & \(<\!0\) \\
    \end{tabular}
  \end{ruledtabular}
\end{table}

Table~\ref{tab:signs_work_heat} summarizes the adopted sign convention for heat and total work across the four operational regimes considered. In this convention, a positive total work \(W>0\) indicates that the system delivers work to the surroundings, whereas \(W<0\) corresponds to work being supplied to the system. Similarly, a positive heat \(Q>0\) denotes energy flowing into the working substance, while \(Q<0\) represents heat released to the external reservoirs. Under these definitions, an engine operates with \(W>0\), absorbing heat from the hot reservoir \((Q_{\mathrm{in}}>0)\) and rejecting heat to the cold one \((Q_{\mathrm{out}}<0)\).

If, under the same parameters, the cycle is run in the reverse (clockwise) direction relative to Fig.~\ref{fig:stirling_engine}, the device operates as a refrigerator. In this case, external work input \((W<0)\) is required to extract heat from the cold reservoir \((Q_{\mathrm{out}}>0)\) and release it to the hot reservoir \((Q_{\mathrm{in}}<0)\).

Beyond these two classical regimes, previous studies have shown that the quantum Stirling cycle may also operate in two additional regimes that arise purely from its quantum character \cite{
WOS:001352772700001,
WOS:001520627100001,
castorene1,araya, 
WOS:001557436100001}. 
In the \textit{accelerator} regime, both the total work and the heat released to the cold reservoir are negative, indicating that external work input is accompanied by simultaneous heating of both reservoirs. In the \textit{heater} regime, all heat flows are negative and \(W<0\), so that the external work is entirely dissipated as heat into both reservoirs. 
Importantly, these operational regimes are not exclusive to intrinsic properties of stacked graphene systems. It has been shown that twisted bilayer graphene, when employed as a Stirling engine at temperatures within 50 to 150 K, can also exhibit this diversity of thermodynamic behaviors \cite{cjlz-lrd6}. 

As will be shown in later sections, the exotic thermodynamic properties of the system, which arise from the constraint of particle number conservation, can induce these additional operational regimes.

\section{Results}
\label{sec:results}

\begin{figure}[htb]
    \centering
    \begin{subfigure}[t]{0.48\textwidth}
        \centering
        \includegraphics[width=\textwidth]{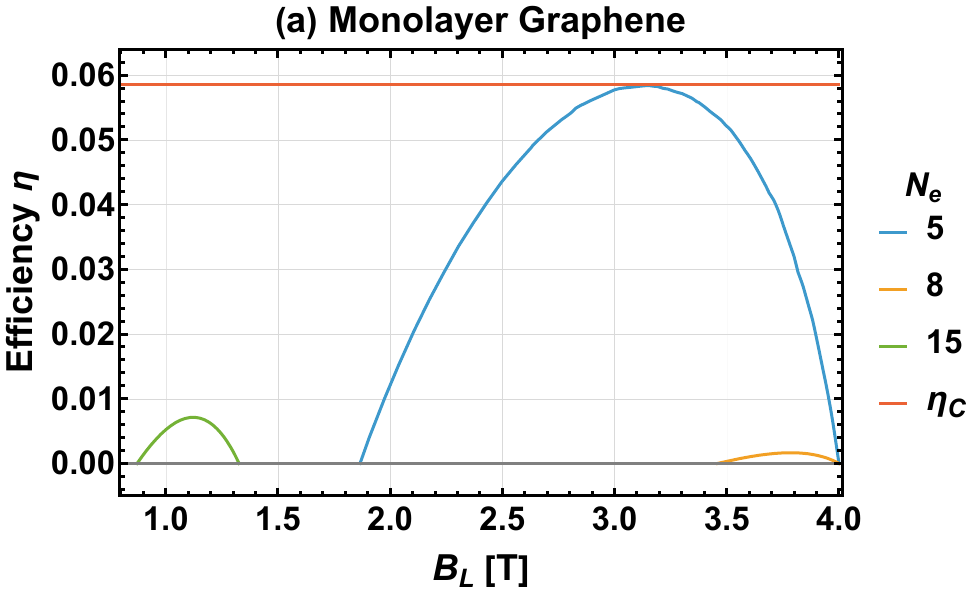}
    \end{subfigure}
    \hfill
    \begin{subfigure}[t]{0.48\textwidth}
        \centering
        \includegraphics[width=\textwidth]{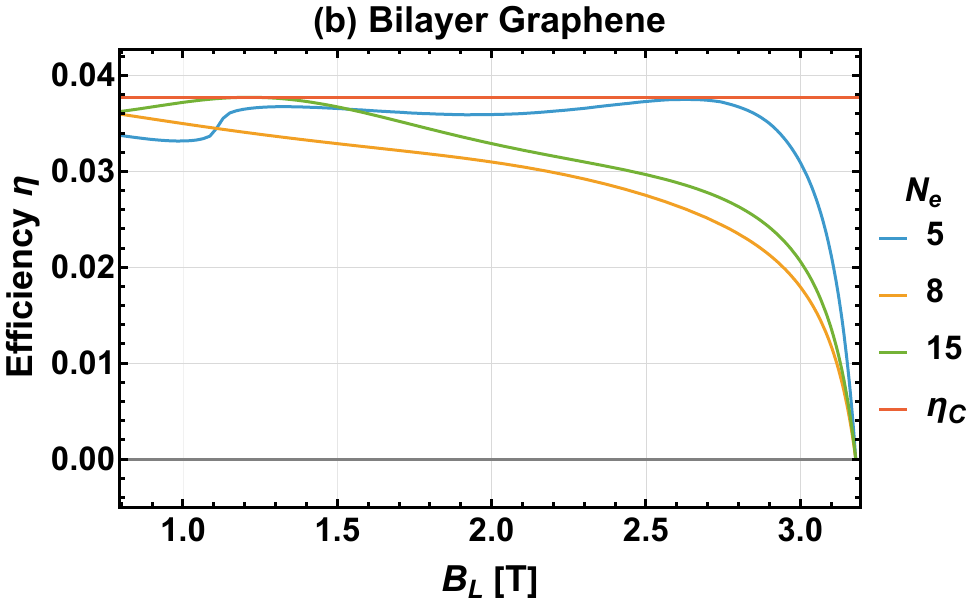}
    \end{subfigure}
    \vspace{0.4em}
    \begin{subfigure}[t]{0.48\textwidth}
        \centering
        \includegraphics[width=\textwidth]{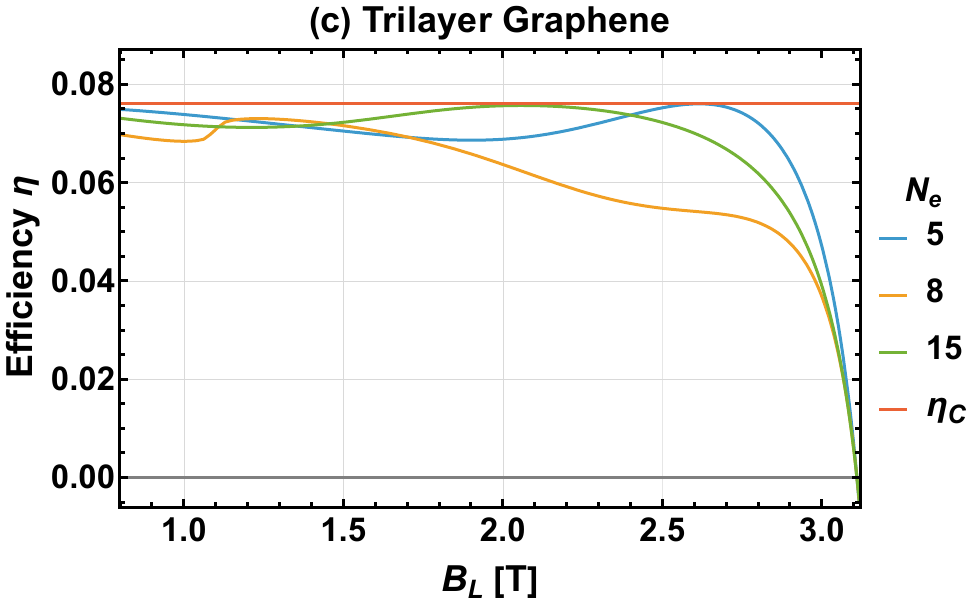}
    \end{subfigure}
\caption{Stirling engine efficiency \(\eta\) as a function of the low magnetic field \(B_L\) for (a) monolayer, (b) bilayer, and (c) trilayer graphene. Each panel shows results for \(N_e = 5, 8, 15\) and the Carnot bound \(\eta_C\). Parameters are: (a) \(T_L = 50.7~\mathrm{K}\), \(T_H = 53.85~\mathrm{K}\), \(B_H = 4.0~\mathrm{T}\), $\eta_C=0.058$; (b) \(T_L = 45.9~\mathrm{K}\), \(T_H = 47.7~\mathrm{K}\), \(B_H = 3.18~\mathrm{T}\), $\eta_C=0.037$; (c) \(T_L = 18.2~\mathrm{K}\), \(T_H = 19.7~\mathrm{K}\), \(B_H = 3.11~\mathrm{T}\), $\eta_C= 0.078$.
}
    \label{fig:stirlingeta}
\end{figure}
\subsection{Stirling Engine Efficiency}\label{sec:efficiency}
In this section, the efficiency of the Stirling engine defined in Eq.~\eqref{efi_eq} and illustrated in Fig.~\ref{fig:stirling_engine} is examined. The working substances correspond to the stacked graphene systems of Fig.~\ref{fig:Diagrama_Grafenos}, with the chemical potential parametrized along all strokes of the cycle to maintain a fixed particle number \(N_e \in \{5,8,15\}\).

To quantify these effects, Fig.~\ref{fig:stirlingeta} presents the efficiency \(\eta\) as a function of the low magnetic field \(B_L\) for monolayer, bilayer, and trilayer graphene. For each configuration, the high magnetic field \(B_H\) and the bath temperatures \(T_L\) and \(T_H\) are kept constant as indicated in the panel captions, while the Carnot efficiency \(\eta_C = 1 - T_L/T_H\) is shown as a horizontal red reference line.
\begin{figure}
    \centering
    \includegraphics[width=1\linewidth]{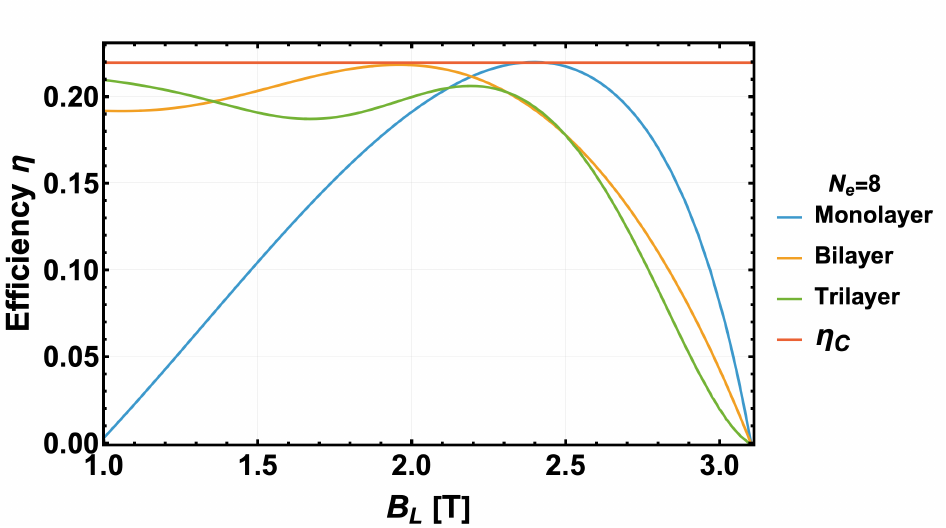}
\caption{Stirling engine efficiency \(\eta\) as a function of the low magnetic field \(B_L\) at fixed high field \(B_H = 3.1~\mathrm{T}\). Reservoir temperatures are set to \(T_L = 21.3~\mathrm{K}\), \(T_H = 27.3~\mathrm{K}\) for the monolayer; \(T_L = 16.7~\mathrm{K}\), \(T_H = 21.4~\mathrm{K}\) for the bilayer; and \(T_L = 10.0~\mathrm{K}\), \(T_H = 12.9~\mathrm{K}\) for the trilayer. In all cases, the Carnot efficiency is \(\eta_C = 0.22\).}
    \label{fig:eficienciasmuchascapas}
\end{figure}

In the monolayer case shown in Fig.~\ref{fig:stirlingeta}(a), the conditions for the system to operate as an engine are strongly constrained for thermal reservoirs at \( T_L = 50.7\,\mathrm{K} \) and \( T_H = 53.85\,\mathrm{K} \), corresponding to a Carnot efficiency of \( \eta_C \simeq 0.06 \), with the high magnetic field fixed at \( B_H = 4.0\,\mathrm{T} \). Although some curves (e.g., \( N_e = 5 \)) can, in principle, achieve the Carnot limit with finite work, this only occurs at low carrier densities and within narrow \( B_L \) intervals. Outside these regions, the device switches to other operational regime modes, and the definition of efficiency is not well-defined. It is for this reason that only small intervals where the substance operates as an engine exist, while no efficiency curves are observed beyond them.  

This restricted behavior can be interpreted as a consequence of the large spacing between relativistic Landau levels, which limits the thermal accessibility of higher-energy states. As a result, the heat exchanged during the isothermal strokes, while retaining the correct sign for engine operation, becomes smaller in magnitude. When the heat contribution from the isomagnetic strokes surpasses that of the isothermal ones—an effect enhanced by the Landau level separation—the net work output turns negative, and the cycle ceases to function as an engine. These exotic behaviors associated with monolayer graphene will be discussed in later sections.

For the bilayer graphene in Fig.~\ref{fig:stirlingeta}(b), the system operates as an engine across the entire \( B_L \) range considered, with the high magnetic field fixed at \( B_H = 3.18\,\mathrm{T} \). The thermal reservoirs are maintained at \( T_L = 45.9\,\mathrm{K} \) and \( T_H = 47.7\,\mathrm{K} \), corresponding to a Carnot efficiency of \( \eta_C \simeq 0.04 \). Under these conditions, the curves for \( N_e = 5 \) and \( N_e = 15 \) attain the Carnot bound at finite values of \( B_L \), with the \( N_e = 15 \) case attaining \( \eta_C \) at lower \( B_L \) values than \( N_e = 5 \). In contrast, the \( N_e = 8 \) curve remains slightly below \( \eta_C \) over the full range, exhibiting a broad maximum just under the Carnot line at $B_L \simeq 0.8 $[T]. This reflects a more balanced competition between the isothermal and isomagnetic contributions arising from the parametrization of the chemical potential. This intermediate behavior highlights how the bilayer spectrum, with its linear scaling in \( B \), provides a broader operational window: depending on the carrier number, the cycle can either saturate the Carnot limit or remain close to it with finite work output. This behavior contrasts with results for twisted bilayer graphene studied within the canonical ensemble \cite{cjlz-lrd6}. While such systems can reach Carnot efficiency in Otto cycles \cite{singh2021magic} and exhibit high efficiencies in quantum Stirling cycles, saturation of the Carnot bound in the Stirling cycle has not been reported in that framework. This suggests that a Fermi–Dirac description with fixed particle number enhances thermodynamic performance in graphene-based quantum heat engines, highlighting the role of ensemble choice and carrier-number constraints in setting efficiency limits.

For the trilayer graphene in Fig.~\ref{fig:stirlingeta}(c), the high magnetic field is fixed at \( B_H = 3.11\,\mathrm{T} \), while the thermal reservoirs are set at \( T_L = 18.2\,\mathrm{K} \) and \( T_H = 19.7\,\mathrm{K} \) with a Carnot efficiency $\eta_C \simeq 0.08$. Under these conditions, the \( N_e = 15 \) curve reaches the Carnot bound near \( B_L \simeq 2.1\,\mathrm{T} \), whereas the \( N_e = 5 \) case attains \( \eta_C \) at \( B_L \simeq 2.65\,\mathrm{T} \), closer to the upper field value \( B_H = 3.18\,\mathrm{T} \). In contrast, the \( N_e = 8 \) curve remains consistently lower across the full range, exhibiting a broad maximum well below \( \eta_C \). This behavior indicates that the enhanced zero-mode degeneracy and the steeper \( B^{3/2} \) scaling of the spectrum restrict the proximity to reversible performance, with the sharpness of the response being induced by the parametrization of the chemical potential. Overall, this smoother yet less favorable trend demonstrates that, although trilayer graphene can reach Carnot efficiency for specific carrier numbers, its operational window in temperature space is narrower than in the bilayer case, and the efficiency maxima are more sensitive to the choice of \( N_e \).
\begin{figure*}[!t]
 \centering
  \begin{subfigure}[b]{0.32\textwidth}
    \includegraphics[width=\textwidth]{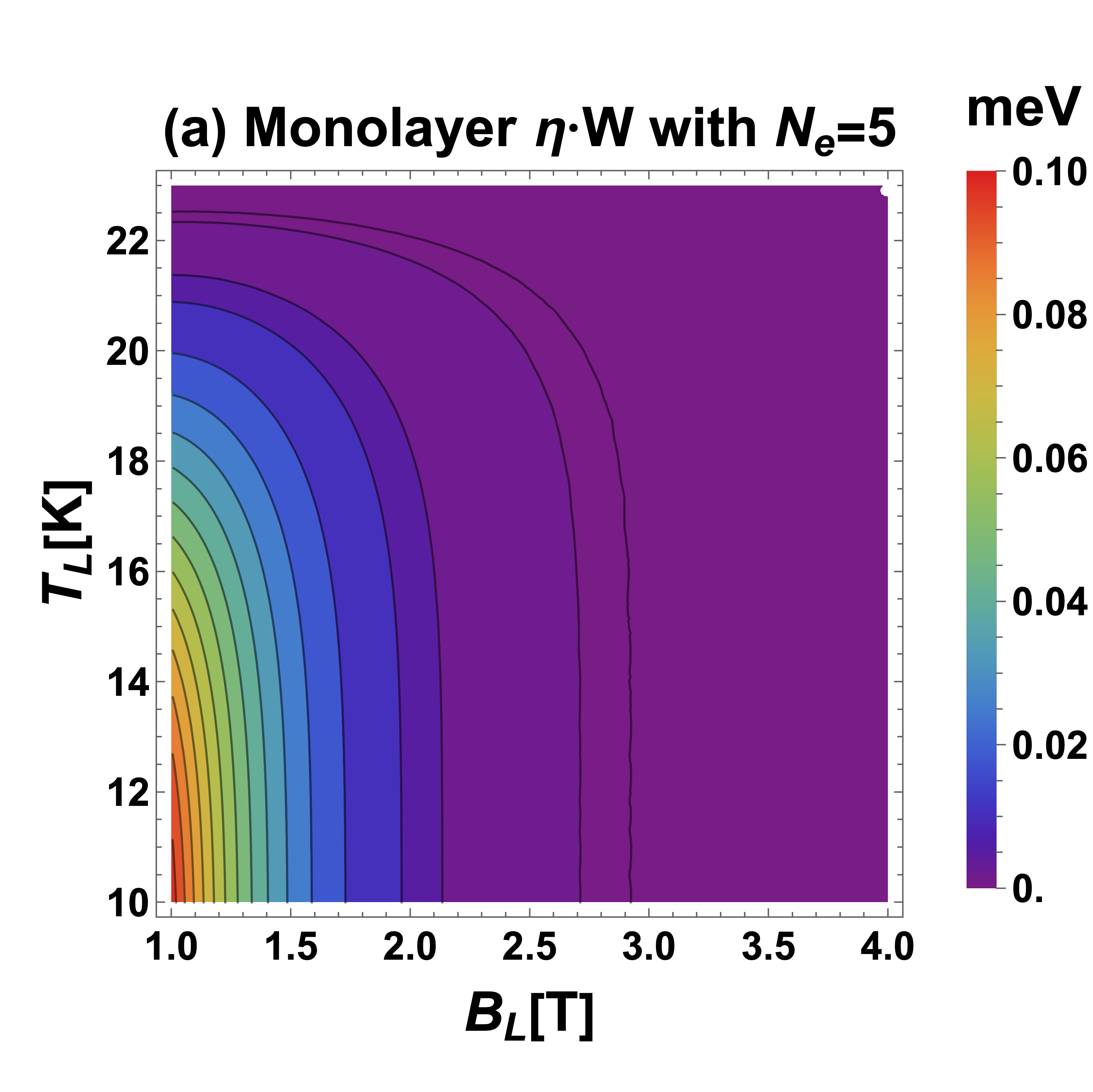}
  \end{subfigure}
  \hfill
  \begin{subfigure}[b]{0.32\textwidth}
    \includegraphics[width=\textwidth]{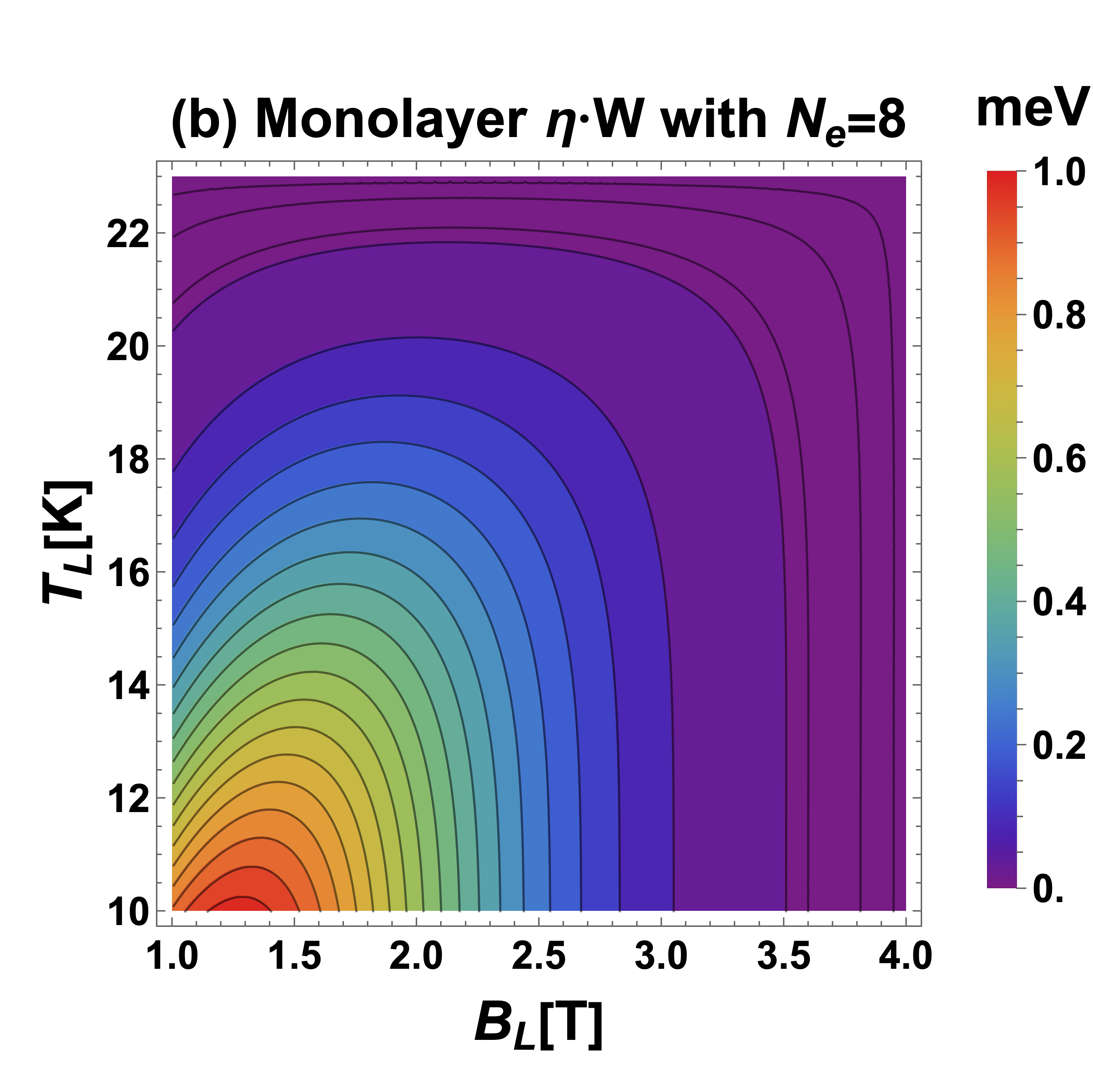}
  \end{subfigure}
  \hfill
  \begin{subfigure}[b]{0.32\textwidth}
    \includegraphics[width=\textwidth]{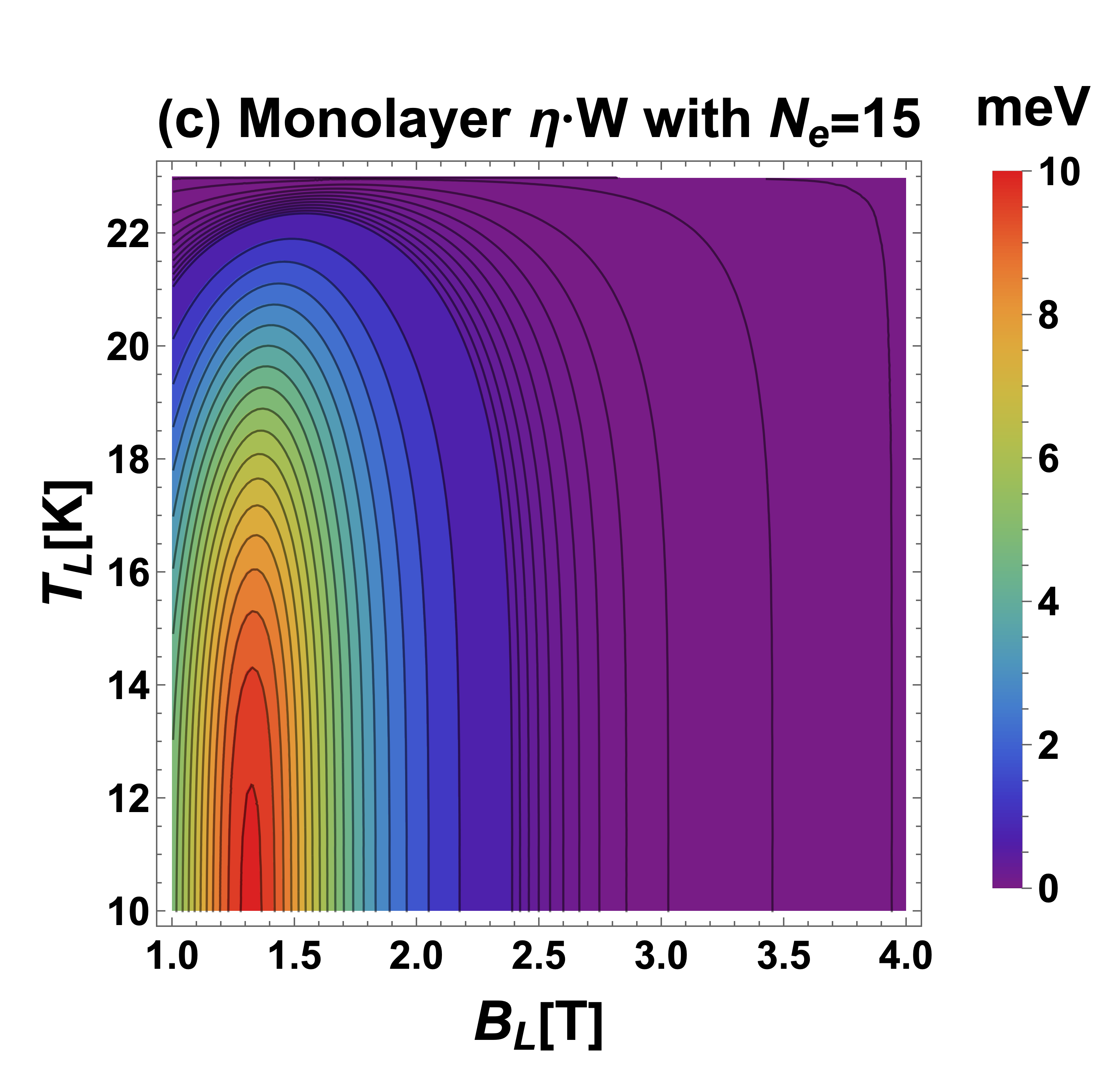}
  \end{subfigure}

  \vspace{0.3cm} 

  \begin{subfigure}[b]{0.32\textwidth}
    \includegraphics[width=\textwidth]{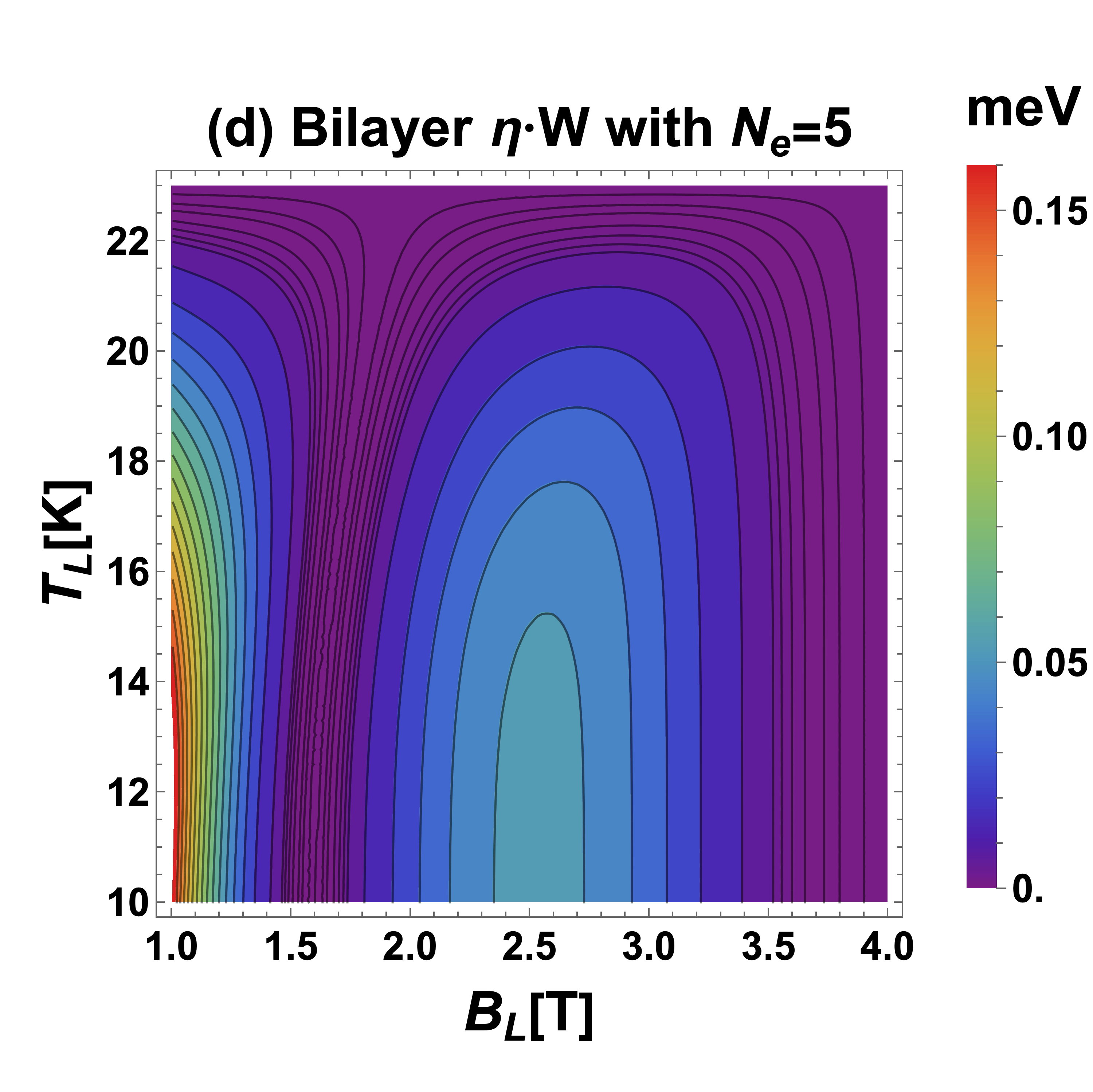}
  \end{subfigure}
  \hfill
  \begin{subfigure}[b]{0.32\textwidth}
    \includegraphics[width=\textwidth]{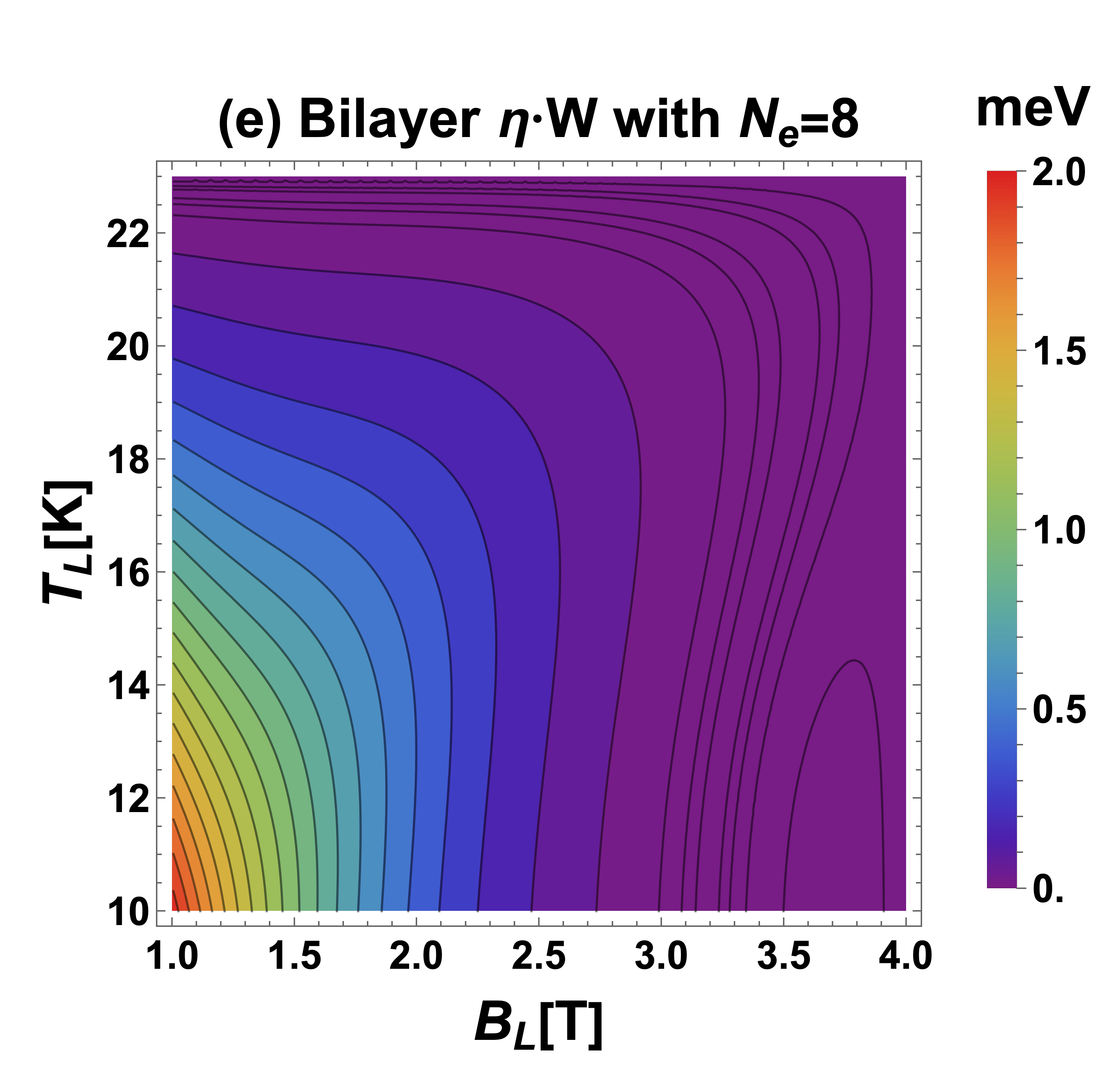}
  \end{subfigure}
  \hfill
  \begin{subfigure}[b]{0.32\textwidth}
    \includegraphics[width=\textwidth]{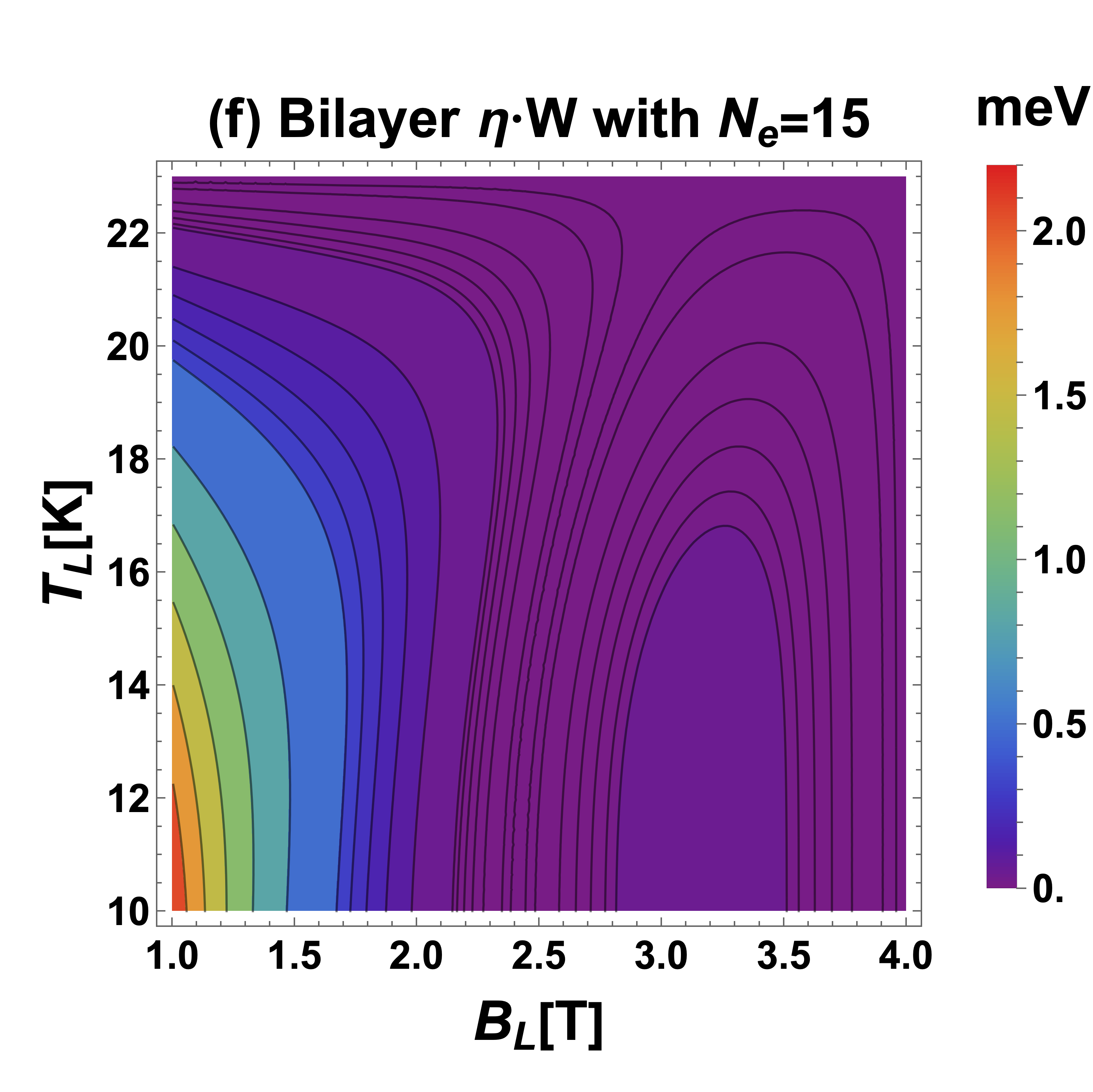}
  \end{subfigure}

  \vspace{0.3cm} 

  \begin{subfigure}[b]{0.32\textwidth}
    \includegraphics[width=\textwidth]{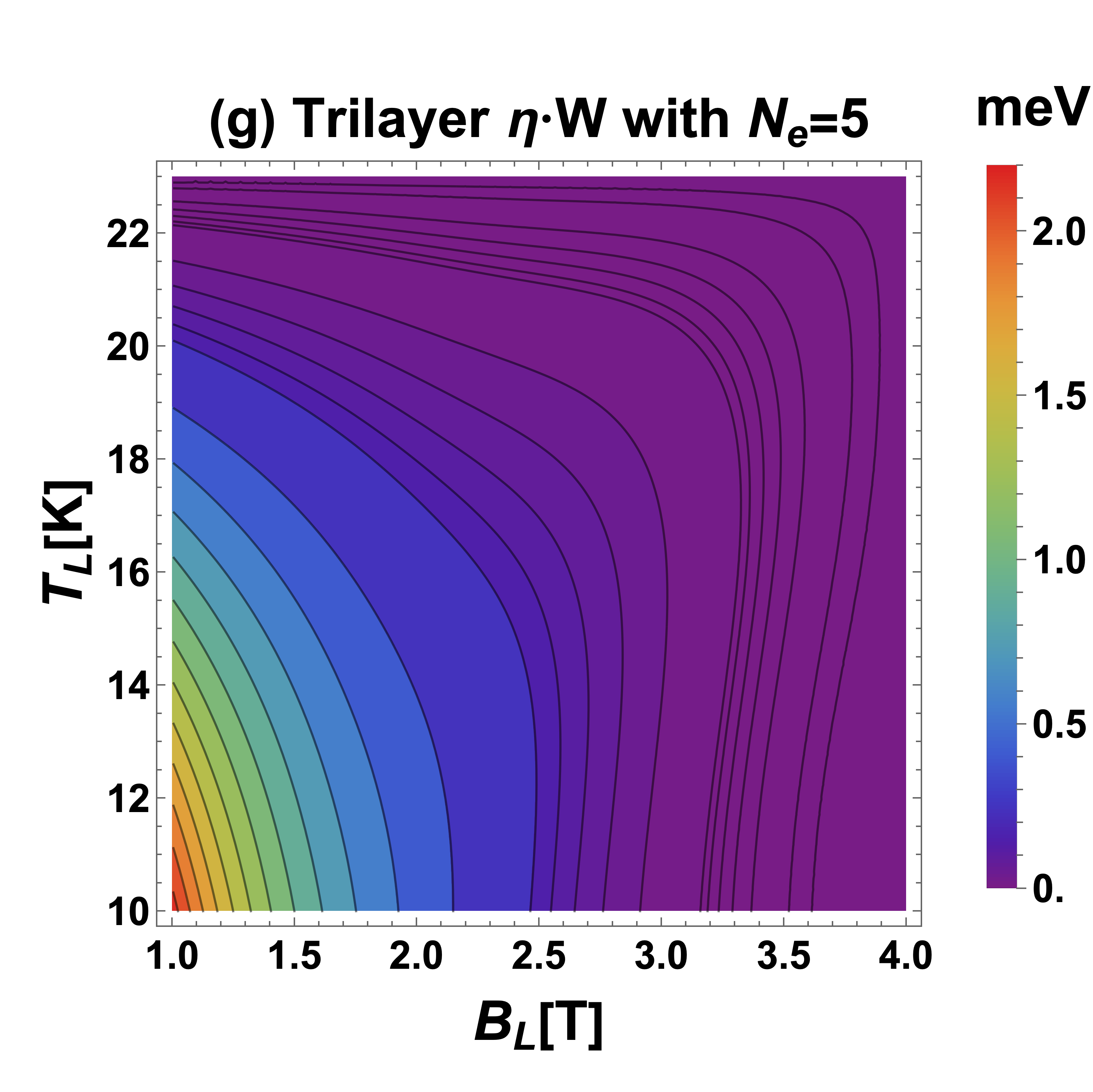}
  \end{subfigure}
  \hfill
  \begin{subfigure}[b]{0.32\textwidth}
    \includegraphics[width=\textwidth]{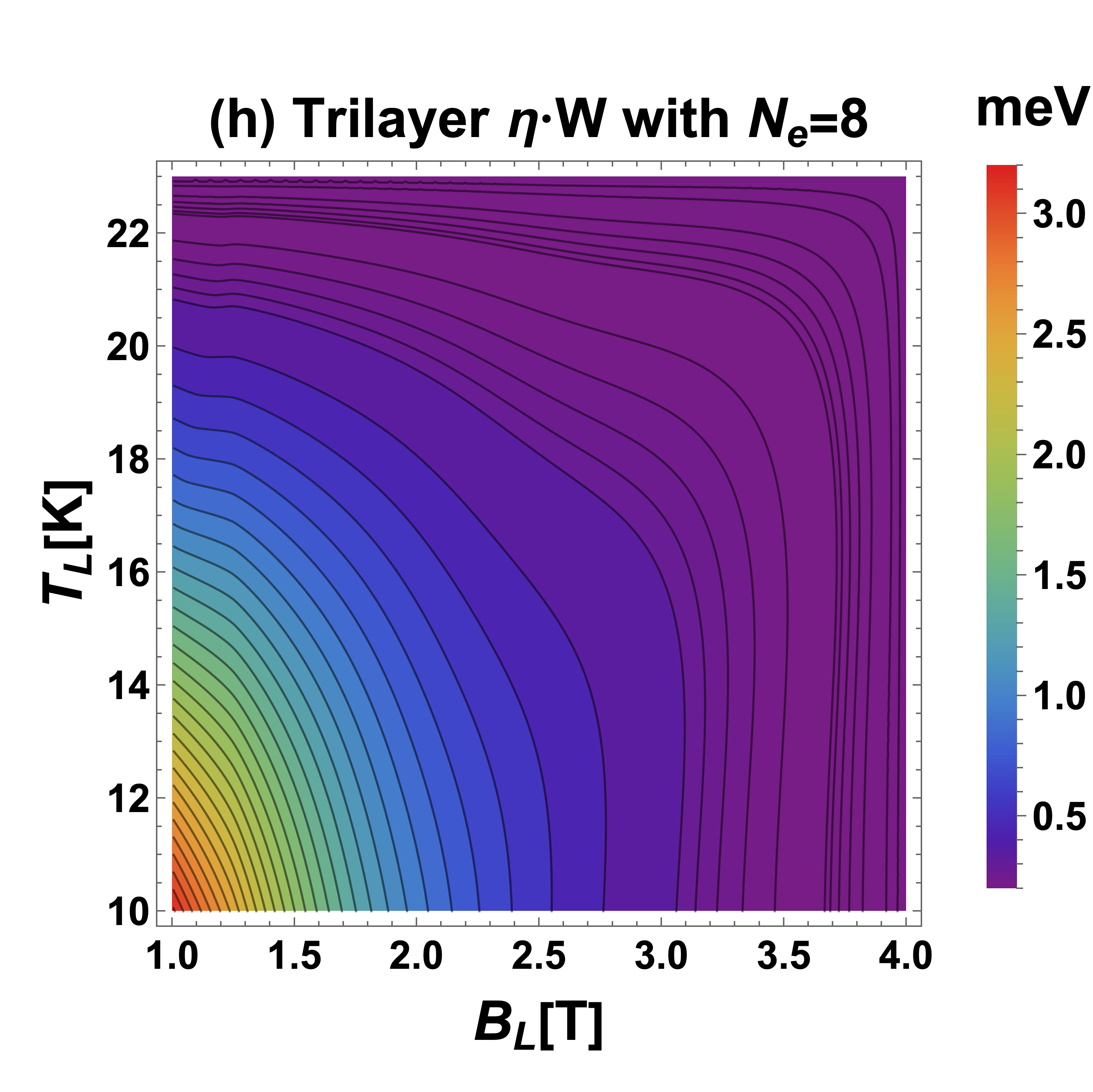}
  \end{subfigure}
  \hfill
  \begin{subfigure}[b]{0.32\textwidth}
    \includegraphics[width=\textwidth]{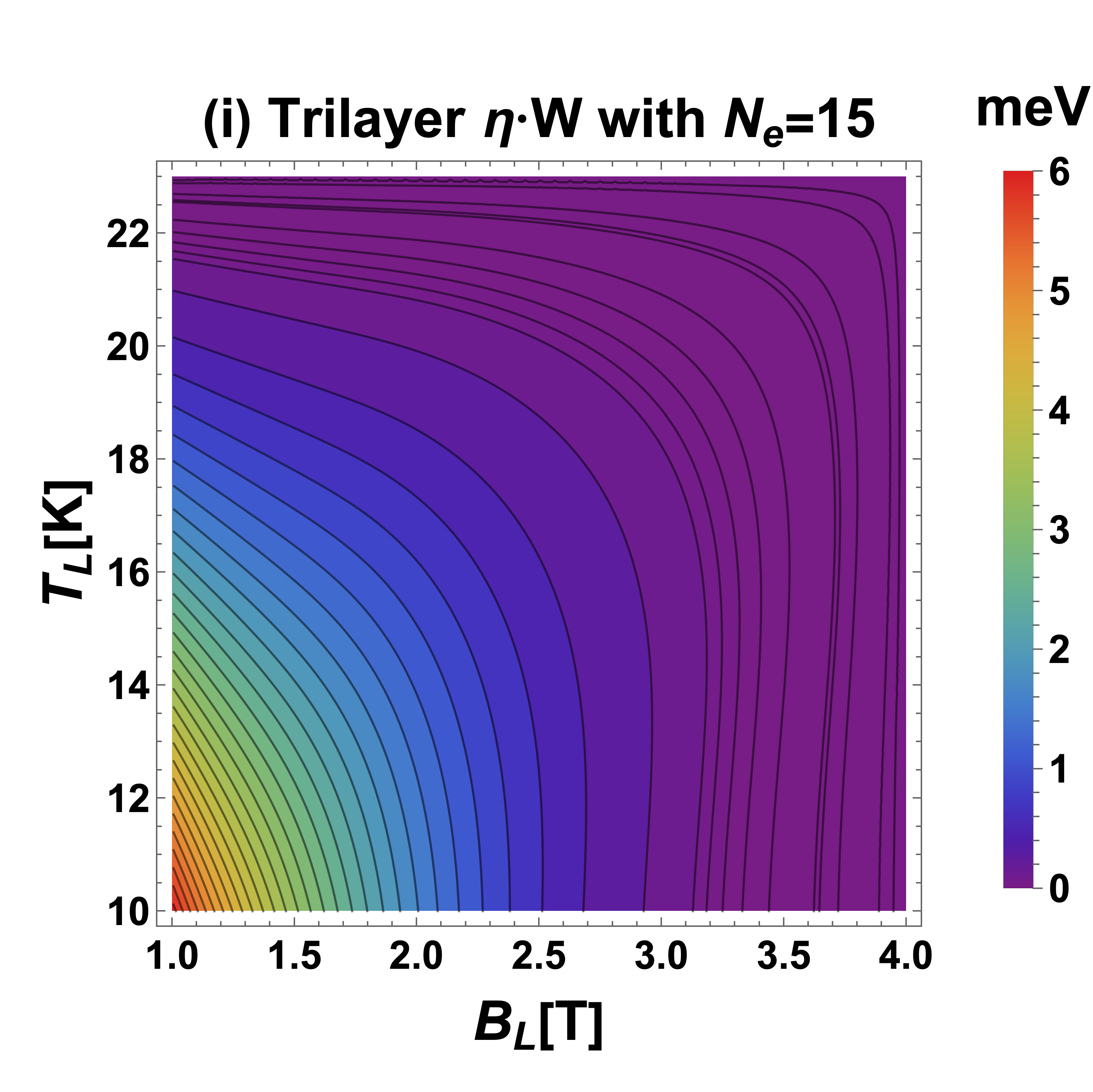}
  \end{subfigure}
    \caption{Contour plots of the product \( \eta \cdot W \) as a function of the low parameters \( B_L \) and \( T_L \), with fixed high parameters \( T_H = 23\,\mathrm{K} \) and \( B_H = 4\,\mathrm{T} \). The panels correspond to monolayer (a)-(c), bilayer (d)-(e) and trilayer (g)-(i) graphene with (left) \( N_e = 5 \), (center) \( N_e = 8 \), and (right) \( N_e = 15 \).}
    \label{fig:etaWmono}
\end{figure*}

 The purpose of Fig.~\ref{fig:stirlingeta} is to demonstrate, over a common range of magnetic field values, that all stacking configurations considered here are capable of reaching the Carnot efficiency. The comparison was designed such that, for each stacking, at least one of the selected particle numbers \( N_e \) attains \( \eta_C \) within approximately the same efficiency window. Because this requires employing similar operational temperatures across the panels, the resulting efficiencies in Fig.~\ref{fig:stirlingeta} appear relatively small. The main motivation behind this choice is to provide a direct and uniform demonstration that the Carnot bound can, in principle, be achieved in all cases, irrespective of the stacking.  

In contrast, Fig.~\ref{fig:eficienciasmuchascapas} focuses on the case \( N_e = 8 \), where all stacking configurations operate as an engine with the same fixed high magnetic field \( B_H = 3.1\,\mathrm{T} \). The thermal reservoirs are chosen such that for the monolayer the temperatures are \( T_L = 21.3\,\mathrm{K} \) and \( T_H = 27.3\,\mathrm{K} \), for the bilayer \( T_L = 16.7\,\mathrm{K} \) and \( T_H = 21.4\,\mathrm{K} \), and for the trilayer \( T_L = 10.0\,\mathrm{K} \) and \( T_H = 12.9\,\mathrm{K} \). Despite these differences, all three cases correspond to the same Carnot efficiency, \( \eta_C \simeq 0.22 \).  

This configuration highlights the role of carrier density in shaping the operational regime. For the trilayer, the maximum efficiency nearly reaches the Carnot limit as \( B_L \to 1.0\,\mathrm{T} \) (the lowest field considered), whereas for the monolayer it occurs at higher fields, around \( B_L \simeq 2.4\,\mathrm{T} \). The bilayer exhibits an intermediate response, achieving the Carnot limit near \( B_L \simeq 1.9\,\mathrm{T} \). Compared to Fig.~\ref{fig:stirlingeta}, which employed a uniform \( B_L \) range to demonstrate that Carnot efficiency can be reached across all stackings, the present figure emphasizes how the optimal field for reversible operation shifts significantly between systems. This contrast underscores the crucial interplay between the Landau-level spectrum and the chemical potential in determining the precise conditions under which Carnot efficiency is realized.

\subsection{Stirling Maximum Gain}

In Sec.~\ref{sec:efficiency}, it was shown that different graphene stackings can reach Carnot efficiency for distinct particle numbers and thermal reservoirs. However, achieving maximum efficiency does not necessarily imply maximizing the extracted work. For this reason, this section analyzes the maximum gain, defined as the product \(\eta \cdot W\) (useful work), as a function of particle number and stacking configuration, while keeping the reservoirs fixed and sweeping over the same magnetic-field range. To illustrate this behavior, Fig.~\ref{fig:etaWmono} displays contour maps of \(\eta W\) as functions of the low parameters \(B_L\) and \(T_L\), with the hot bath fixed at \(T_H = 23~\mathrm{K}\) and \(B_H = 4.0~\mathrm{T}\). The rows correspond to monolayer, bilayer, and trilayer graphene, while the columns show the cases \(N_e \in \{5,8,15\}\). In all configurations, the best performance is concentrated at low magnetic fields (\(B_L \lesssim 1.5~\mathrm{T}\)) and moderately low temperatures (\(T_L \sim 10{-}15~\mathrm{K}\)), reflecting the enhancement of work and efficiency in regimes where a denser set of thermally accessible Landau levels is available.  

In the monolayer case, shown in Figs.~\ref{fig:etaWmono}(a)–(c), the high-\(\eta W\) region is narrow and quickly suppressed as either \(B_L\) or \(T_L\) increase. The maximum for \(N_e = 5\) occurs at the lowest field and within \(T_L \in 10{-}13\,\mathrm{K}\). For \(N_e = 8\), the favorable region expands slightly in field range (\(B_L \in 1.0{-}1.5\,\mathrm{T}\)) but narrows in temperature (\(T_L \in 10{-}11\,\mathrm{K}\)). At \(N_e = 15\), the contour shrinks again in field range relative to \(N_e = 8\), though it widens in temperature (\(T_L \in 10{-}14\,\mathrm{K}\)). The maximum values also increase markedly with particle number, from about \(0.1\,\mathrm{meV}\) at \(N_e=5\) to \(1.0\,\mathrm{meV}\) at \(N_e=8\), and nearly \(10\,\mathrm{meV}\) at \(N_e=15\).  

For bilayer graphene, illustrated in Figs.~\ref{fig:etaWmono}(d)–(f), the optimum broadens into a smoother, rounded peak, indicating a wider operational window consistent with the enhanced density of states. For \(N_e=5\), the maximum again appears near the lowest field but over a broader temperature range (\(T_L \simeq 10{-}15\,\mathrm{K}\)), together with a secondary plateau around \(B_L \simeq 2.5\,\mathrm{T}\). For \(N_e=8\), the maximum shifts back to very low fields and a narrow window of \(T_L \simeq 10{-}11\,\mathrm{K}\). At \(N_e=15\), the contours resemble the previous case but display a low-\(\eta W\) pocket at \(B_L \simeq 2.8{-}3.4\,\mathrm{T}\). The scaling with particle number differs from the monolayer: the maxima are \(0.15\,\mathrm{meV}\) for \(N_e=5\), \(2.0\,\mathrm{meV}\) for \(N_e=8\), but remain at \(2.0\,\mathrm{meV}\) for \(N_e=15\), showing saturation rather than continued growth.  

The trilayer configuration, depicted in Figs.~\ref{fig:etaWmono}(g)–(i), sustains sizable values of \(\eta W\) over a comparable \(B_L\) range and extends to slightly higher \(T_L\). At \(N_e=5\), the maximum is located near the lowest field and around \(T_L \simeq 10\,\mathrm{K}\), reaching \(2.0\,\mathrm{meV}\). For \(N_e=8\), the contours broaden both in field and temperature, with the maximum increasing to \(3.0\,\mathrm{meV}\). At \(N_e=15\), the pattern closely resembles the \(N_e=8\) case but with a higher maximum of about \(6.0\,\mathrm{meV}\), still significantly smaller than the monolayer for the same particle number. This reduction is a direct consequence of the smaller Landau-level spacing and lower energy scale in the trilayer system.

\subsection{Stirling Operational Regimes}
\begin{figure}
    \centering
    \includegraphics[width=.93\linewidth]{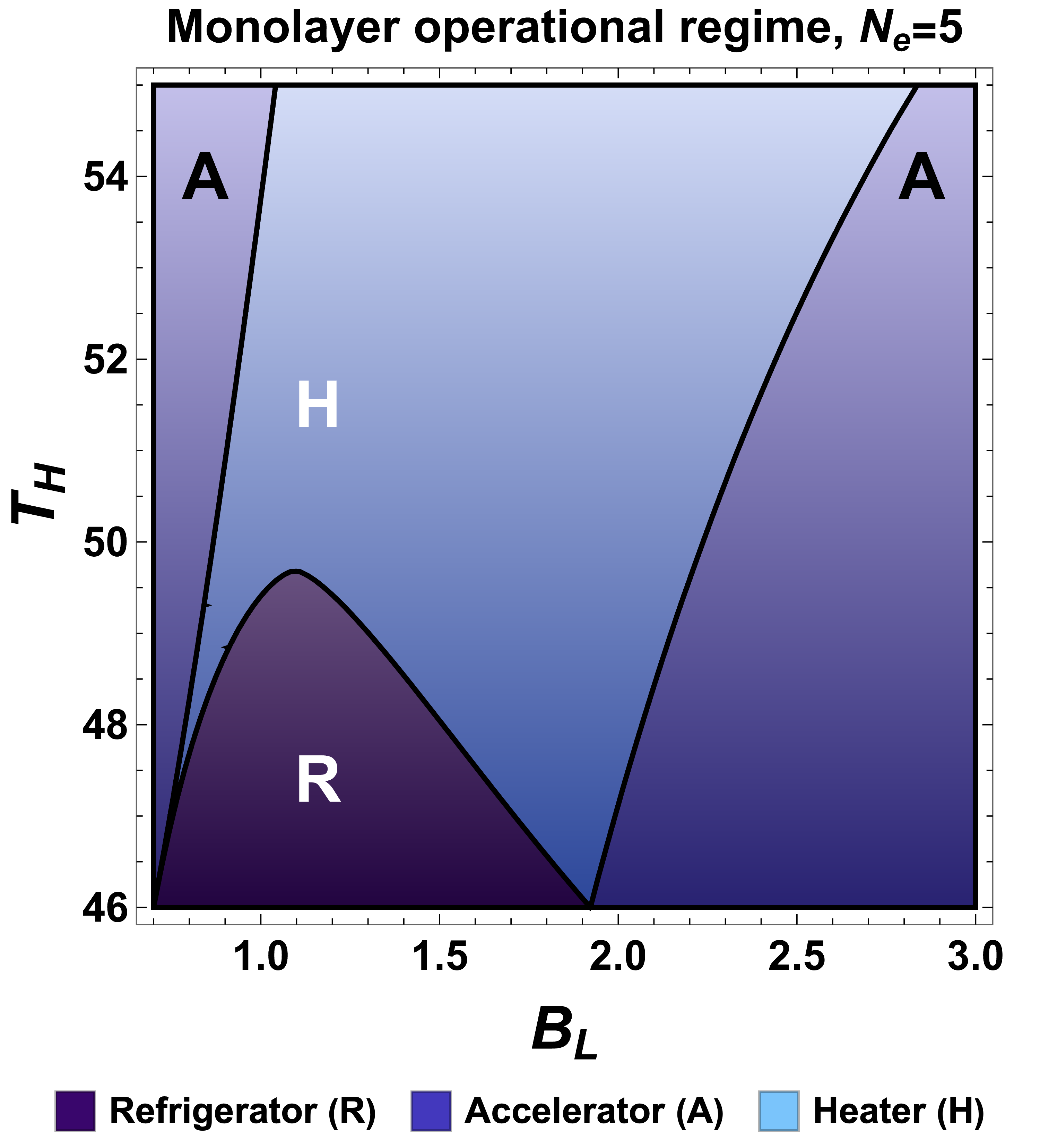}
\caption{Operational regimes of the Stirling cycle for monolayer graphene with \(N_e = 5\), as functions of the low magnetic field \(B_L\) and hot reservoir temperature \(T_H\), at fixed \(B_H = 3.0~\mathrm{T}\) and \(T_L = 46~\mathrm{K}\).}
    \label{fig:regimenes}
\end{figure}
In the previous sections, the working substances were shown to operate as engines with high efficiency and significant work extraction. Nevertheless, these systems may also exhibit exotic behavior, leading to the emergence of distinct operational regimes. To illustrate this point, Fig.~\ref{fig:regimenes} presents the operational map of the Stirling cycle for monolayer graphene at fixed \(B_H = 3.0~\mathrm{T}\), \(T_L = 46~\mathrm{K}\), and \(N_e = 5\). Within the explored window of low magnetic field \(B_L\) and hot reservoir temperature \(T_H\), no engine region is observed. Instead, the diagram reveals a small refrigerator domain at \(B_L \simeq 0.7{-}1.8~\mathrm{T}\) and \(T_H \simeq 46{-}49.5~\mathrm{K}\), two accelerator wedges near the lowest and highest values of \(B_L\), and a broad heater region covering most of the parameter space.  

The absence of engine operation indicates that, under these conditions, heat exchange in the isomagnetic strokes dominates over the isothermal contributions. This outcome reflects both the wide spacing of relativistic Landau levels and the parametrized chemical potential imposed to maintain \(N_e = 5\) throughout the cycle. Consequently, the system exhibits \(W < 0\) across the entire diagram, with the specific regime determined by the balance of heat flows: refrigerator, accelerator, or heater. These results align with the performance maps in Fig.~\ref{fig:etaWmono}, where the monolayer shows weak \(\eta W\) values and a rapidly vanishing optimum, as well as with the efficiency curves in Fig.~\ref{fig:stirlingeta}, which display only small, low-efficiency engine regions and predominantly non-engine behavior. Engine windows may emerge under different parameter choices, but are absent in the present configuration.

This does not exclude bilayer graphene from exhibiting the full range of operational regimes. Previous studies have shown that the same Stirling cycle implemented in bilayer graphene can display all operational regimes under different parameter choices and when the chemical potential is not fixed \cite{cjlz-lrd6}. Their absence here therefore reflects the specific thermodynamic window considered, rather than an intrinsic limitation of the system.

It should be emphasized that realizing different operational regimes in bilayer and trilayer graphene would require substantially lower temperatures and much higher magnetic fields than those explored in this study, owing to their smaller characteristic energy scales. Accordingly, the present analysis focuses on regimes accessible at higher temperatures, which are not only more relevant to this work but also more feasible from an experimental perspective.

At the same time, it is important to note that the present analysis is intended
as a theoretical proof of concept rather than a concrete experimental proposal.
While the main text focuses on a fixed-particle-number formulation in order to
enable a controlled comparison between different multilayer systems, an
alternative and experimentally natural realization of the Stirling cycle
consists in fixing the chemical potential. This extension, which mimics the
situation of a graphene system coupled to electronic reservoirs, is explicitly
addressed in Appendix~\ref{app:fixed_mu}, where we show that the qualitative
features of the operational regimes remain robust with respect to the choice of
thermodynamic ensemble.

In a realistic implementation, field-driven modifications of the Landau--level
structure would induce variations of the sample magnetization across the
different strokes of the cycle. When the graphene system is
electromagnetically coupled to an external circuit, these magnetization changes
can generate an electromotive response, allowing the conversion of
field-induced energy variations into usable electrical work. This mechanism
provides a physically consistent route for work extraction without requiring
commitment to a specific device architecture.

\section{Conclusions and discussion}
\label{sec:conclusion}

This study investigates quantum Stirling cycles in monolayer, AB-stacked bilayer, and ABC-stacked trilayer graphene subjected to perpendicular magnetic fields.
Using a grand-canonical, fully fermionic framework based on exact Landau spectra, we demonstrate that the cycles can operate within experimentally accessible regimes. 
This is supported by an order-of-magnitude comparison between simulated particle numbers, chemical potentials, and sheet carrier densities defined by the magnetic length.  

Thermodynamic quantities, including internal energy and entropy, were consistently evaluated using Fermi–Dirac statistics, with the chemical potential \( \mu (T,B)\) parametrized to conserve particle number along both isothermal and isomagnetic branches. In this framework, the perpendicular magnetic field acted as the external control parameter driving the cycle.  

Across the explored parameter space and for multiple fixed particle numbers, the monolayer and the stacked configurations demonstrated the capacity to reach Carnot efficiency under suitable magnetic conditions. Among them, the AB bilayer exhibited the broadest operational window and reached the Carnot bound while sustaining finite work output, whereas the monolayer showed more constrained operation and the trilayer presented smoother efficiency trends with sizable values of the combined performance metric \(\eta W\) over a comparable \(B_L\) range.  

Overall, the monolayer Stirling cycle, due to its larger Landau-level spacing and higher characteristic energy scale, delivered greater useful work at higher efficiency compared with the bilayer and trilayer cases. Its capability to operate across all thermodynamic regimes as an engine, refrigerator, heater, and accelerator highlights its versatility. This feature suggests possible applications, such as employing the monolayer cycle to cool other working substances including quantum batteries or supercondensates, while simultaneously acting as a heater at elevated temperatures.

{Our comparative analysis shows that the efficiency and work output of a
Stirling cycle are strongly influenced by the stacking-dependent Landau–level structure of graphene multilayers. 
This demonstrates that simply varying the layer configuration introduces distinct thermodynamic behaviors and efficiency profiles, proving that novel effects can be engineered without altering the intrinsic material composition. 
Remarkably, certain parameter regimes allow the cycle to approach Carnot efficiency while retaining finite
work, a behavior that emerges solely from the intrinsic spectral properties of each stacking. 
These results highlight stacking order as a meaningful design knob for tailoring the thermodynamic response of quantum heat engines.}
 
\section*{Data availability statement}
The data that support the findings of this article are openly available at Zenodo \cite{castorene_2026_19389102}.

\begin{acknowledgments} 
B.C, F.J.P, M.H.G, N.C, and P.V. acknowledge financial support from ANID
Fondecyt grant no. 1240582. B.C, F.J.P., M.H.G, and P.V. acknowledge financial support from ANID
Fondecyt grant no. 1250173. B.C acknowledge PUCV and Programa de Incentivo a la Iniciación Científica (PIIC) no. 004 from ''Direcci\'on de Postgrado'' of UTFSM. 
B.C. acknowledges the support of ANID Becas/Doctorado Nacional 21250015.
C.L. thanks the support by the Brazilian agencies CNPq and FAPERJ.
M.H.G acknowledge PUCV and ''Direcci\'on de Postgrado'' of UTFSM.
B.C, F.J.P, M.H.G and P.V acknowledge partial support CEDENNA Grant CIA25002.
\end{acknowledgments}

\appendix

\section{Stirling cycle at fixed chemical potential}
\label{app:fixed_mu}

In the main text, the Stirling cycle was primarily formulated under a
fixed-particle-number constraint, with the chemical potential determined
self-consistently as $\mu=\mu(T,B)$. An alternative and experimentally natural formulation consists in fixing the chemical potential throughout the cycle, corresponding to a graphene system coupled to an electronic reservoir. In this Appendix, we present additional results obtained within this fixed-$\mu$ formulation in order to assess the robustness of the conclusions reported in the main text.

\subsection{Efficiency and work at fixed \texorpdfstring{$\mu$}{mu}}

\begin{figure}[t]
    \centering
    \includegraphics[width=1.0\columnwidth]{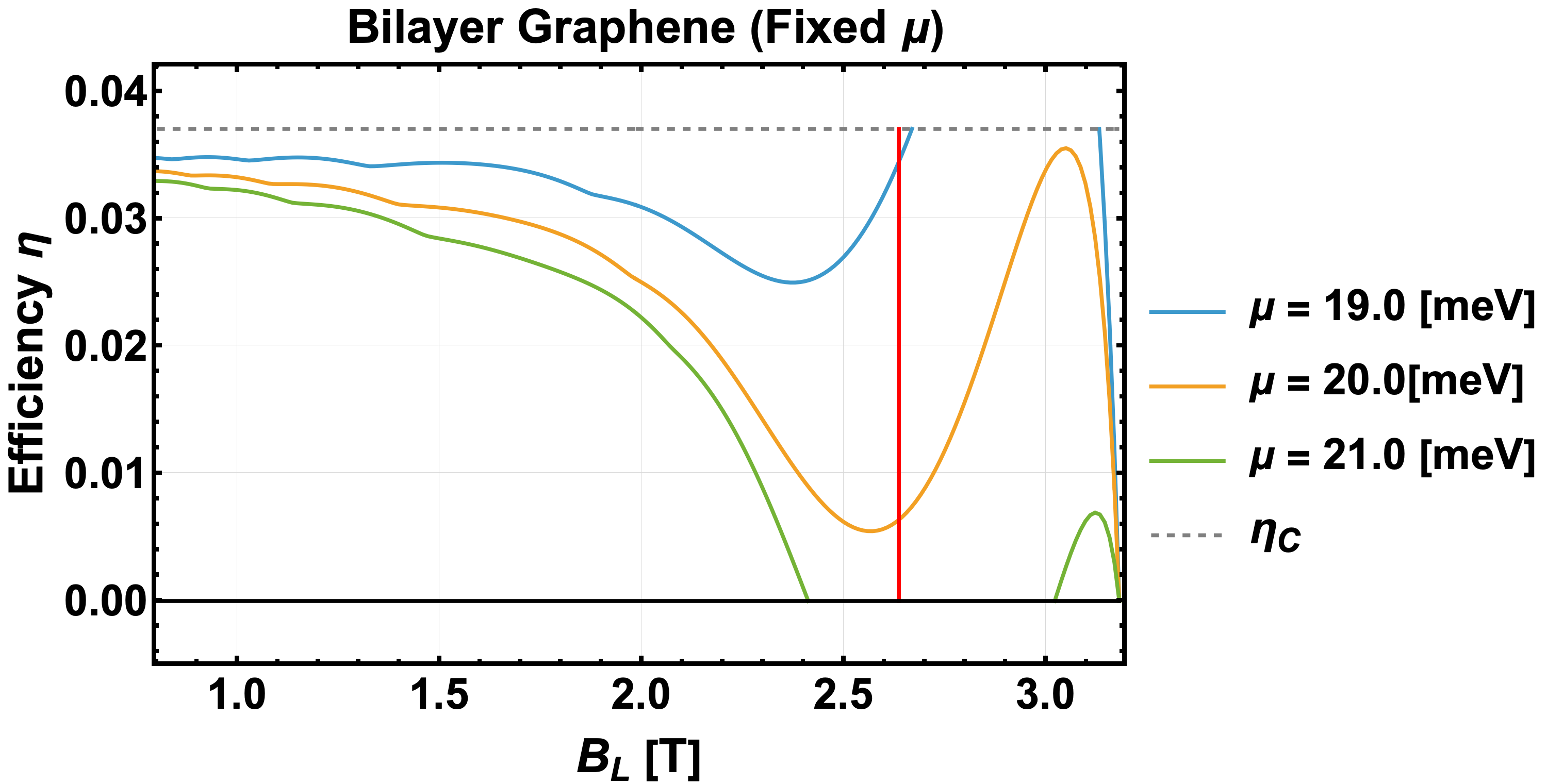}
    \caption{Stirling-cycle efficiency $\eta$ of AB-stacked bilayer graphene as a
    function of the lower magnetic field $B_L$, computed at fixed chemical
    potential $\mu=19$, $20$, and $21~\mathrm{meV}$. The cycle parameters are the
    same as those used in the main text:
    $T_L = 45.9~\mathrm{K}$, $T_H = 47.7~\mathrm{K}$, and
    $B_H = 3.18~\mathrm{T}$, corresponding to a Carnot efficiency
    $\eta_C = 0.037$ (horizontal dashed line). The vertical red line marks the
    value of $B_L$ at which the fixed-particle-number protocol with $N_e=5$
    yields $\eta=\eta_C$ in the main analysis. {For values exceeding $\eta_C$ or falling below zero, the system transitions to a different operational regime.}}
    \label{fig:mu_fixed_bilayer}
\end{figure}

We first analyze the Stirling-cycle efficiency and total work per cycle for
AB-stacked bilayer graphene by keeping the chemical potential fixed during the
entire cycle. The bilayer system is chosen here because it exhibits the broadest
operational window in the fixed-$N_e$ analysis discussed in the main text.

Fig.~\ref{fig:mu_fixed_bilayer} shows the efficiency $\eta$ as a function of
the lower magnetic field $B_L$ for several representative values of $\mu$, using
the same cycle parameters as in the main text,
$T_L = 45.9~\mathrm{K}$, $T_H = 47.7~\mathrm{K}$, and $B_H = 3.18~\mathrm{T}$, for
which the Carnot efficiency is $\eta_C = 0.037$. The vertical red line indicates
the value of $B_L$ at which the fixed-particle-number protocol with $N_e=5$
attains $\eta=\eta_C$. {For values below zero and above $\eta_{C}$, the definition of efficiency is no longer applicable because the system transitions to a different operational regime. To avoid possible confusion, these regions are therefore not plotted.}

\begin{figure}[t]
    \centering
    \includegraphics[width=1.0\columnwidth]{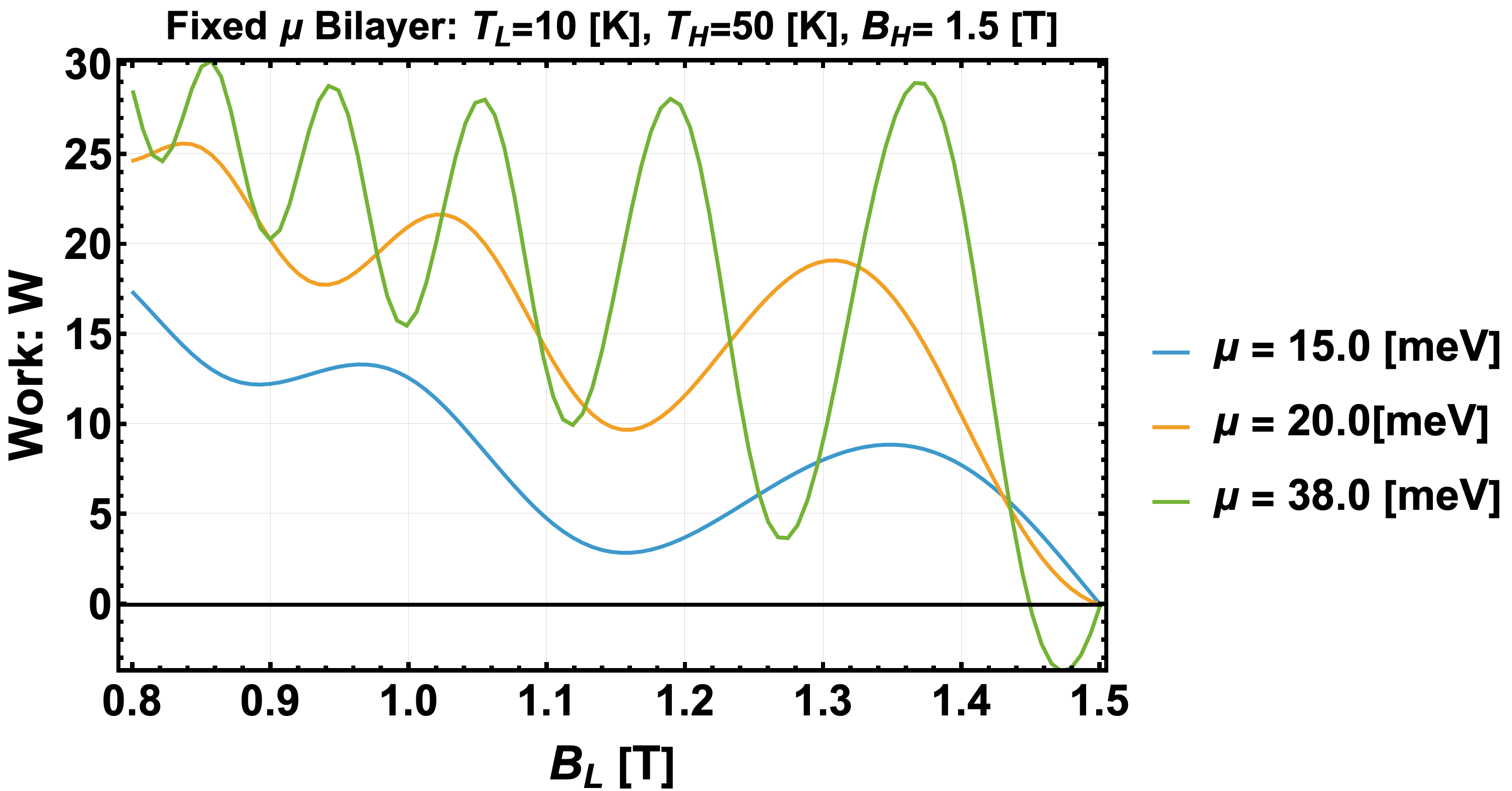}
    \caption{Total work per cycle $W$ for AB-stacked bilayer graphene as a
    function of the lower magnetic field $B_L$, computed at fixed chemical
    potential $\mu=15$, $20$, and $38~\mathrm{meV}$. The cycle parameters are
    $T_L=10~\mathrm{K}$, $T_H=50~\mathrm{K}$, and $B_H=1.5~\mathrm{T}$. The oscillatory behavior originates from successive 
{crossings of the chemical potential through the Landau levels},
which modulate the electronic population and the heat exchanged during the
isothermal strokes. {For value falling below zero, the system transitions to a different operational regime.}}
    \label{fig:work_fixed_mu_bilayer}
\end{figure}

\begin{figure}[h!]
    \centering
    \includegraphics[width=1.0\columnwidth]{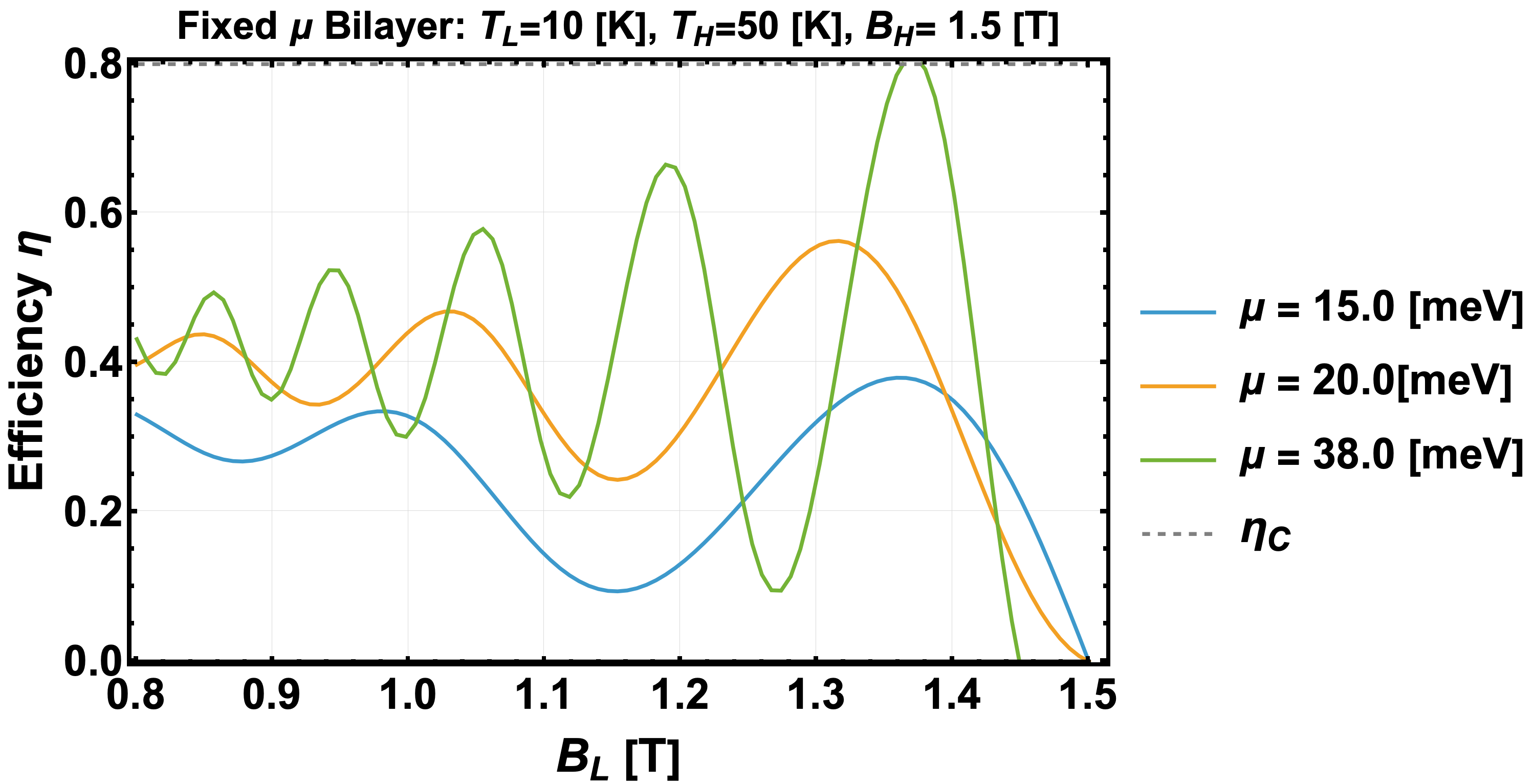}
    \caption{Stirling-cycle efficiency $\eta$ of AB-stacked bilayer graphene as a
    function of the lower magnetic field $B_L$, computed at fixed chemical
    potential $\mu=15$, $20$, and $38~\mathrm{meV}$, using the same cycle
    parameters as in Fig.~\ref{fig:work_fixed_mu_bilayer}. The horizontal dashed
    line indicates the Carnot efficiency $\eta_C$. Despite the oscillatory features induced by 
{the crossings of the chemical potential through the Landau levels}, 
well-defined high-efficiency operating regimes persist. {For values falling $\eta_C$ below zero, the system transitions to a different operational regime.}}
    \label{fig:eff_fixed_mu_bilayer}
\end{figure}

At fixed $\mu$, both the efficiency and the extracted work exhibit oscillatory
structures as functions of $B_L$, as shown in
Figs.~\ref{fig:work_fixed_mu_bilayer} and \ref{fig:eff_fixed_mu_bilayer}. These oscillations originate from successive
Landau-level crossings through the fixed chemical potential, which modulate the
electronic population and, consequently, the heat exchanged during the
isothermal strokes. Despite these quantitative differences with respect to the
fixed-$N_e$ formulation, well-defined operating windows with sizable work output
and high efficiency persist. This confirms that the qualitative behavior of the
bilayer Stirling engine is not an artifact of the ensemble choice.

\subsection{Particle-number variation at fixed \texorpdfstring{$\mu$}{mu}}

\begin{figure}[t]
    \centering
    \includegraphics[width=0.75\columnwidth]{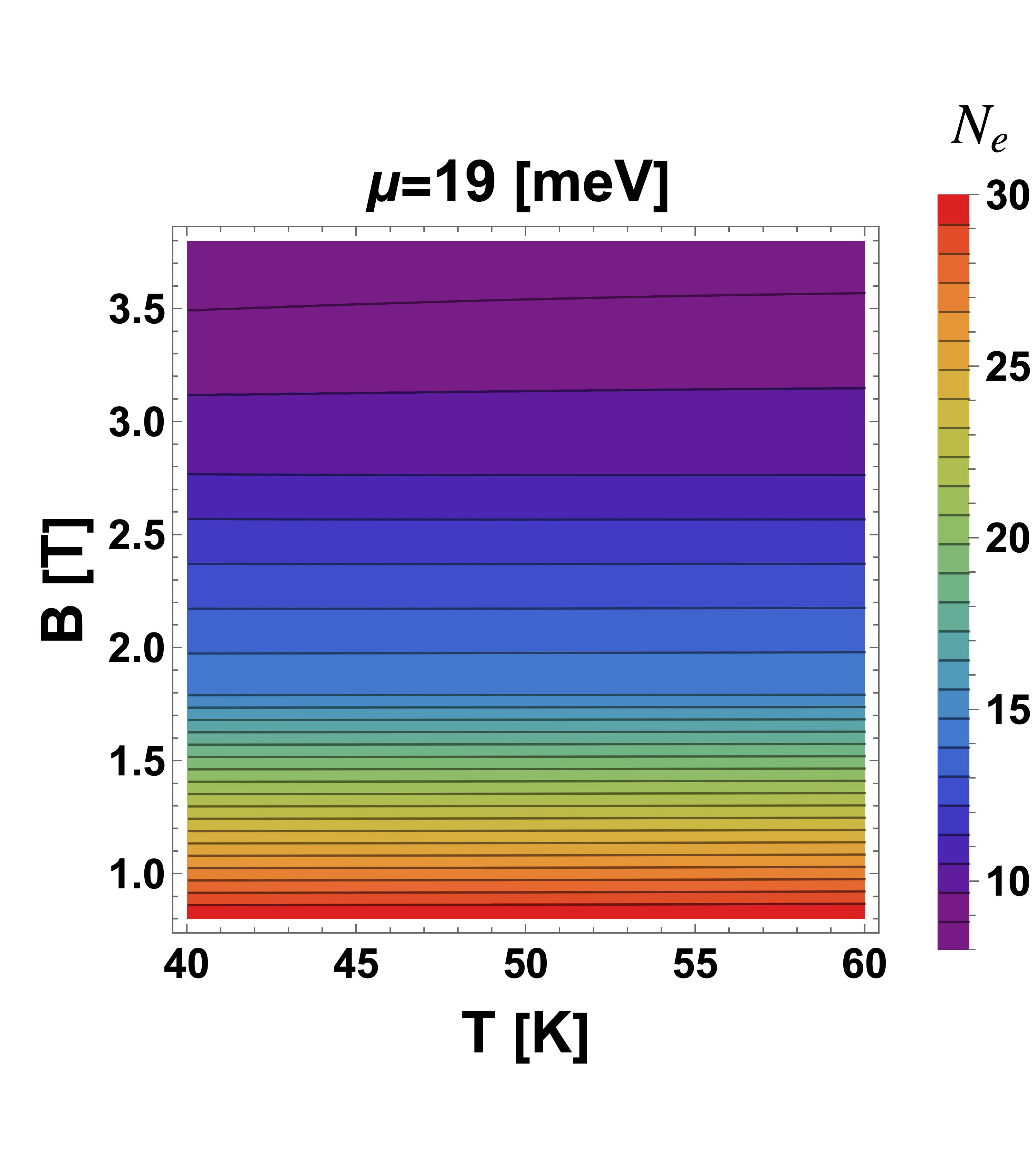}
    \caption{Contour plot of the particle number $N_e$ as a function
    of temperature $T$ and magnetic field $B$ for AB-stacked bilayer graphene at
    fixed chemical potential $\mu=19~\mathrm{meV}$. The parameter range
    corresponds to that used in the main text. In this region, the nearly horizontal contour lines
    indicate that the particle number varies only weakly with temperature,
    explaining the consistency between the fixed-$N_e$ and fixed-$\mu$
    formulations in the operating regime of interest.}
    \label{fig:mu_fixed_N_contour}
\end{figure}

To further clarify the relation between the fixed-$N_e$ and fixed-$\mu$
formulations, Fig.~\ref{fig:mu_fixed_N_contour} presents a contour plot of the
particle number $N_e$ as a function of temperature $T$ and magnetic
field $B$ at fixed chemical potential $\mu=19~\mathrm{meV}$. 
The parameters correspond to the same range used in the main text. 

In this parameter range, the particle number exhibits only weak variations with
temperature, as indicated by the nearly horizontal contour lines. This shows
that, close to the operating point relevant for the Stirling cycle, fixing
$\mu$ effectively corresponds to an approximately constant particle number.
This explains why the fixed-$N_e$ and fixed-$\mu$ formulations yield consistent
results in the regime explored in the main text.

Overall, the analysis presented in this Appendix demonstrates that the
performance and operational regimes of the bilayer graphene Stirling engine are
robust with respect to the choice of thermodynamic ensemble, and that the main
conclusions of the manuscript are rooted in the underlying Landau-level
structure rather than in a particular constraint on $\mu$ or $N_e$.

%


\end{document}